\RequirePackage[2020-02-02]{latexrelease}
\documentclass[aps,prl,reprint,amssymb,groupedaddress,twocolumn]{revtex4}
\usepackage{graphicx} 
\usepackage{subfigure}
\usepackage{hyperref,hypcap}
\usepackage{braket}
\usepackage{amsmath}
\usepackage{multirow}
\usepackage{bm}

\begin{document}
\title{Exact Diagonalization for Magic-Angle Twisted Bilayer Graphene}
\author{Pawel Potasz}
\affiliation{Department of Physics, The University of Texas at Austin, Austin, Texas 78712, USA}
\affiliation{Department of Physics, Wroclaw University of Science and Technology, 50-370 Wroclaw, Poland} 
\author{Ming Xie}
\author{A. H. MacDonald}
\affiliation{Department of Physics, The University of Texas at Austin, Austin,  Texas 78712, USA}

\date{\today}

\begin{abstract}
We report on finite-size exact-diagonalization calculations in a Hilbert space defined by 
the continuum-model flat moir\'e bands of magic angle twisted bilayer graphene (MATBG).
For moir\'e band filling $3>|\nu|>2$, where superconductivity is strongest,
we obtain evidence that the ground state is a spin ferromagnet.
Near $|\nu|=3$, we find Chern insulator ground states that have spontaneous spin, valley, and sublattice 
polarization, and demonstrate that the anisotropy energy in this order-parameter space is 
strongly band-filling-factor dependent.  We emphasize that inclusion of the
remote band self-energy is necessary for a reliable description of MATBG flat band correlations. 
\end{abstract}

\maketitle

{\em Introduction:}---
Near a magic twist angle, the width of bilayer graphene's low energy moir\'e bands
shrinks \cite{MorellFlatBand, BMModel} by an order of magnitude or more, allowing interactions to 
play a prominent role in shaping electronic properties.
The flat bands form an octet that is the direct product of two-fold spin, valley, and band or 
sublattice degrees of freedom and closely analogous to the spin/valley/layer octet of Bernal bilayer graphene \cite{KharitonovMaxim, ZhangNiuMacDonald, MacDonaldJungZhang, BarlasMacDonald}.
The recent discovery of superconductivity and interaction-induced Chern 
and trivial insulator states \cite{Tutuc2017,CaoInsulator,CaoSuper,YoungDean,MacDonaldEfetov, UriJarilloHerrero, StepanovEfetov, BalentsYoung, WongYazdani, ZondinerIlani, SharpeGoldhaber, SerlinYoung, WuAndrei, AroraNadjiPerge, DasEfetov, ParkFlavor, CaoNematicity, StepanovCompeting, RozenIlani} in magic-angle  twisted bilayer graphene (MATBG) has motivated ongoing theoretical work \cite{FuModel, VishwanathOrigin, Scalettar, Phillips, SpinLiquid, WangModel, MottAF, Yang, PALee,Vafek, Rademaker, KoshinoFu, Bascones, Ochi, FuJune, AFMonHoneyComb, XuBalents, FuSuper, Juricic, Das, Super1, Super2, Super3, Super4, GuineaInteractions, MingHF, SantosEarly, VishwanathWannierObstructions, VishwanathFaithful, TarnopolskyVishwanath, LiuDai, Semenoff, ZhangChern, Jung2014, SaitoYoung, AjeshCollectiveModel, TBGI, TBGII, TBGIII, TBGIV, TBGV, TBGVI, MingHF2, BultinckZaletel, BultinckZaletel2, ChatterjeeZaletel, ChatterjeeZaletel2, KhalafVishwanath, KhalafVishwanathZaletel, ParkerZaletelBultinck,KangVafek,PizarroWehling,KangVafek2,RepellinSenthil, LiaoQMC, SoejimaParkerZaletel,KangVafek3}, from which 
it is already clear that, although MATBG states share properties with doped and undoped Mott
insulators in conventional crystals, they also have a relationship to integer 
and fractional quantum Hall (FQH) states \cite{VishwanathWannierObstructions, VishwanathFaithful, BultinckZaletel}.

Progress in understanding competitions between different low energy states and the 
sensitivity of the ground state properties to particular model parameters has been achieved using numerical mean-field theory \cite{WangModel,MingHF,MingHF2,MacDonaldEfetov, StepanovCompeting,BultinckZaletel2,KhalafVishwanathZaletel,ParkerZaletelBultinck,AjeshCollectiveModel}, and beyond, using exact diagonalization \cite{Ochi,RepellinSenthil,TBGVI}, quantum Monte Carlo \cite{MottAF,PALee,LiaoQMC} and 
density matrix renormalization group methods \cite{KangVafek3, RepellinSenthil, ParkerZaletelBultinck, SoejimaParkerZaletel,ChatterjeeZaletel2},
and using both Hubbard-like lattice \cite{WongYazdani,ZondinerIlani,RozenIlani,FuModel,WangModel,Ochi,KangVafek2,MottAF,Yang,PALee,LiaoQMC,Vafek,KoshinoFu,VishwanathOrigin,VishwanathWannierObstructions,VishwanathFaithful,Bascones,AFMonHoneyComb,XuBalents} and continuum models \cite{StepanovCompeting,MingHF,MingHF2,MacDonaldEfetov,BultinckZaletel2,KhalafVishwanathZaletel,RepellinSenthil, ParkerZaletelBultinck, AjeshCollectiveModel,TBGI, TBGII, TBGIII, TBGIV, TBGV, TBGVI,SoejimaParkerZaletel,KangVafek3}. In this Letter we use exact diagonalization to describe correlations within flat bands 
that are identified by solving the single-particle problem \cite{BMModel} exactly.  The use of numerical flat bands in place of approximate 
Wannier orbitals has the advantage that we account accurately for crucial changes in the charge distribution
of flat band wave functions as a function of moir\'e Brillouin-zone momentum. 
We use a systematic approach that accounts fully for self-energies from remote bands, which
play a key role, to make further progress.  Because the MATBG octet enlarges finite Hilbert space sizes
far beyond those of spinful single-band models, we are forced to restrict our attention primarily 
to flat band filling factors with $|\nu| \geq 2$; fortunately much of the strong correlation physics seen experimentally
occurs in this filling factor regime.

\begin{figure}[ht]
\begin{center}
\includegraphics[width=0.95\columnwidth]{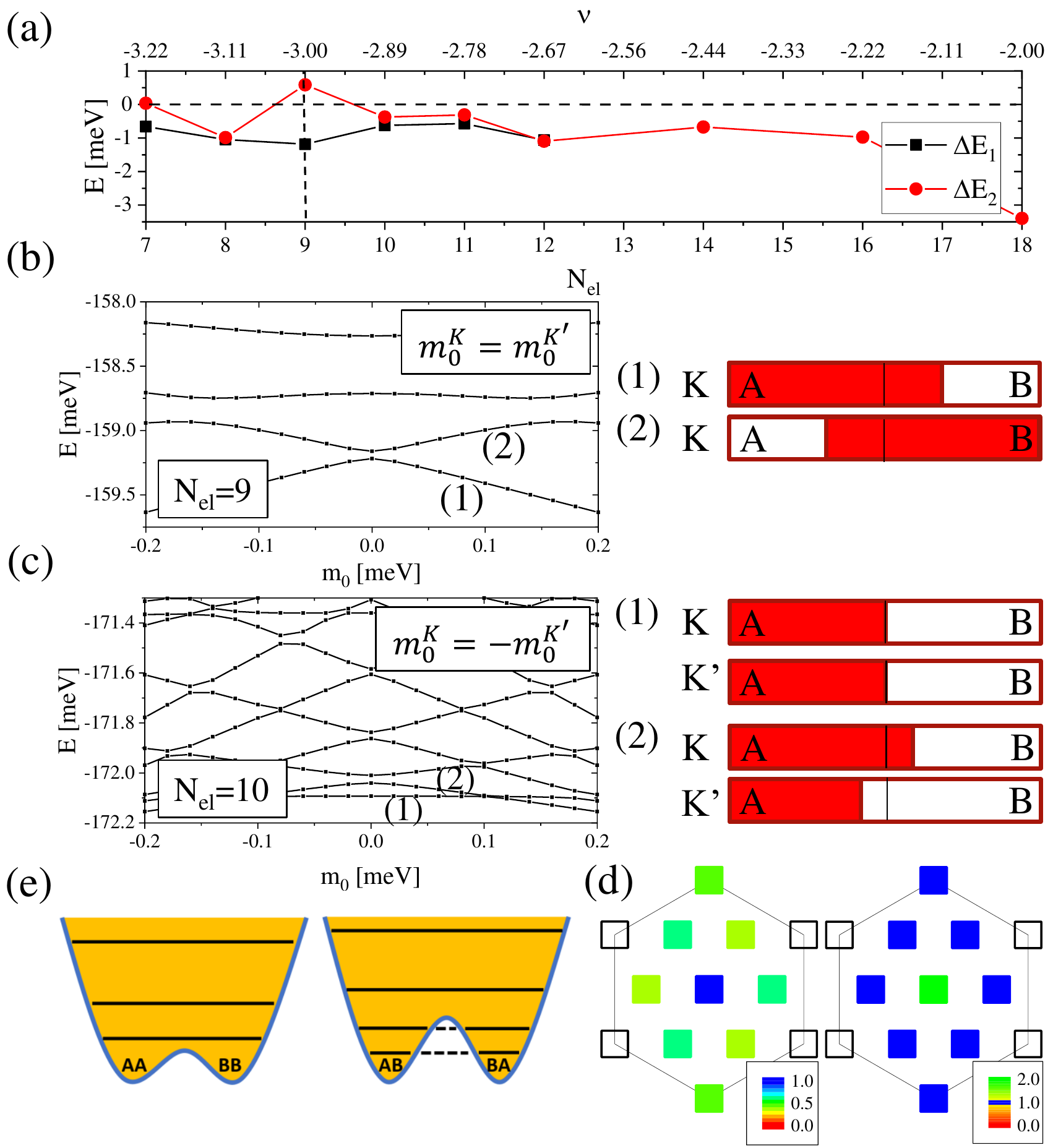}
\caption{Spin, valley, and sublattice order {\it vs.} electron number $N_{el}$
in finite-size MATBG with $M=9$ moir\'e unit cells:
(a) $\Delta E_{1} = E_{min}(S_{max}) - E_{min}(S<S_{max})$ (meV/unit cell), 
where $S = S_K + S_{K'}$ is total spin and $S_{max}=N_{el}/2$ is its maximal value.
$\Delta E_{2} =E_{min}(P_v=0)-E_{min}(P_v=1)$ where $P_v = |N_K-N_{K'}|/(N_K+N_{K'})$.
(b) Response of the ground state energy of the valley polarized state at $|\nu|=3$ to an external field that couples to sublattice polarization. (c) Response of low energy states to a valley-odd sublattice field at $N_{el}=10$. 
The right panels in (b) and (c) schematically illustrate the sublattice polarizations induced 
in states (1) and (2) by the corresponding fields. (d) Ground state momentum space occupation numbers projected to valley $K$ (left), and traced over valley (right) at $N_{el}=10$.
(e)  Schematic illustration of macroscopic quantum tunneling of the sublattice pseudospin
collective coordinate, based on $N_{el}=10$ ED results.
The ground state (1)(left) has the same sense of
sublattice polarization in the two valleys, and at this system size, strong hybridization between the  
sublattice polarization states labeled $AA$ and $BB$, where the first letter corresponds to valley $K$ and the second to 
valley $K'$.
The first excited state (2) (right) has opposite sense of sublattice polarization in opposite valleys ($AB$) and ($BA$), 
and much weaker hybridization between the two degenerate states at this system size.  
}   
\label{Fig:Fig1}
\end{center}
\end{figure}

Our calculations confirm \cite{DasEfetov, WuAndrei, ParkFlavor, StepanovCompeting}
that spin, valley, and sublattice polarization is common in both insulating 
and metallic states, demonstrate that the anisotropy energy associated 
with these generalized ferromagnetic orders is strongly filling factor dependent, and provide evidence for  
spin-polarized ground states for $|\nu|\in (2,3)$ - the range of filling factor that 
supports the strongest superconductivity.
This picture is revealed in exact diagonalization (ED) finite-size system results by signatures of macroscopic quantum tunneling.
Our main results are presented in Fig.~\ref{Fig:Fig1} where 
panel (a) provides evidence that ground states are maximally spin-polarizated for $|\nu|\in (2,3)$,
but valley-polarized only near $|\nu|=3$.  Figure \ref{Fig:Fig1}(b) shows that the ground state at $|\nu|=3$ 
is a spin and valley polarized doublet formed by states with opposite senses of spontaneous sublattice
polarization. These states are known to be 
Chern insulators and are accurately approximated by Hartree-Fock theory.
The ground state of the system with one charge added to 
(or removed from) the $|\nu|=3$ ground state (Fig.~\ref{Fig:Fig1}(c)) is still fully spin polarized,
but completely loses its $K,K'$ valley polarization.  
As shown in Fig.~\ref{Fig:Fig1}(d) these states nevertheless have precisely integer 
occupation numbers for all momenta, but only when summed over valleys.  We conclude 
that the states with added and removed charge have easy-plane valley order; we attribute the sudden change in 
anisotropy to the strong band/sublattice dependence of the single-particle Hamiltonian at 
momenta near the $\gamma$ point in the moir\'e Brillouin zone.
The sublattice polarization-properties (Fig.~\ref{Fig:Fig1}(e)) of the ground states near $|\nu|=3$,
discussed further below,
are revealed by the responses to sublattice and valley dependent potentials
illustrated in Figs.~\ref{Fig:Fig1}(b) and \ref{Fig:Fig1}(c).

{\em Flat band projected exact diagonalization:}---
Because of large Dirac velocities, the electronic 
density of states of an isolated neutral graphene sheet 
has a minimum at neutrality and is small over a broad energy range, 
allowing interaction effects to be described perturbatively.
When magic-angle moir\'e bands \cite{BMModel} are formed, strong electronic correlations 
emerge and perturbative analyses are less reliable. 
The ED of the Hamiltonian is a powerful nonperturbative method to study strong correlations,
but, because the many-body Hilbert space grows exponentially with system size, it
is practical only when the single-particle Hamiltonian can be truncated to a reasonably small dimension,
typically with at most several tens of single-particle states. 
In MATBG the spectral isolation of the eight flat bands of interest 
(flat conduction and valence bands for each of four spin or valley flavors) 
motivates projection to an occupation number subspace in which all remote valence bands in graphene's negative-energy 
sea are fully occupied, all remote conduction bands are 
empty, and occupation numbers are allowed to fluctuate only within the flat bands. 
This strategy leads to a low-energy effective Hamiltonian that acts entirely in the flat-band Hilbert space:
\begin{eqnarray}
H_{eff}&=& \sum_{i',i} \; [ \epsilon_{i} \delta_{i',i}
+ \Sigma_{i',i} ] \, c^{\dagger}_{i'} c_{i} \nonumber 
\\ &+& \frac{1}{2} \sum_{i',i,j',j} \langle i', j' |V|i,j\rangle \; c^{\dagger}_{i'} c^{\dagger}_{j'} c_{j} c_{i},
\label{Eq:Heff}
\end{eqnarray} 
where $ \langle i', j' |V|i,j\rangle$ is a two-body Coulomb interaction matrix element, $i',i,j',j$ label flat band states, $\epsilon_{i}$ is an eigenvalue of the single-particle twisted-bilayer graphene
Hamiltonian \cite{BMModel} including the interlayer tunneling contribution that is responsible for flat band formation, and 
\begin{eqnarray} 
\Sigma_{i',i} &=& \sum_{v} \; [ \langle i', v |V|i,v \rangle - \langle i', v |V|v,i \rangle] \nonumber \\
&-& \sum_{\bar{v}} \; [ \langle i', \bar{v} |V|i,\bar{v}\rangle - \langle i', \bar{v} |V|\bar{v} ,i \rangle ],
\label{Eq:RemoteBandSE2}
\end{eqnarray}
which we refer to the remote band self energy, 
accounts for Hartree and exchange interactions with states $v$ in the frozen negative energy sea.
In Eq.~(\ref{Eq:RemoteBandSE2}) the sum over $\bar{v}$ in the regularization term is over the
frozen valence bands of a neutral bilayer with no-interlayer tunneling \cite{MingHF}.
As we shall emphasize, the remote band self-energy plays an essential role in MATBG 
physics and unlike in the related case of Landau level physics, cannot be neglected. Its importance derives from the 
fact that flat valence band wave functions have strongly 
momentum-dependent spatial distributions across the moir\'e unit cell, even when 
averaged over the full band \cite{Rademaker,GuineaInteractions}.  
This issue is solved by appropriately renormalizing the flat bands by adding 
self-energies from the remote valence bands.  Both Hartree and Fock terms are essential when considering the physics away from the neutrality point (fully filled flat valence band) in effective Hamiltonians projected to flat band subspace.  
This self-energy accounts for leading-order interactions between flat and 
remote bands, and includes exchange interactions that enhance intersubband layer coupling 
as emphasized in a recent perturbative renormalization group calculation by Kang and Vafek \cite{KangVafek}. 
At higher order, remote band polarization will screen the Coulomb interaction in Eq.~(\ref{Eq:Heff}),
among other less understood effects. We partially account for these screening effect \cite{PizarroWehling} by allowing the 
(in general $q$ dependent) dielectric function used in constructing the Coulomb matrix elements to
be larger than the value that would be expected on the basis of dielectric and gate screening alone.

The remote band self-energy reshapes the bands principally by shifting 
energies near $\gamma$ upward, relative to 
those near $\kappa$, $\kappa'$.  The relative shifts occur primarily because the 
Hartree potential from the remote bands is attractive near the $AA$ positions where states near $\gamma$
have less weight \cite{Rademaker, GuineaInteractions, MingHF}.
The sharp contrast between the conduction and valence band widths in these 
empty-band dispersions does not imply strong 
particle-hole asymmetry.  Indeed the model we will study is very nearly  
particle-hole symmetric, and the relative widths of the bands is reversed 
when we describe flat band states in terms of interacting holes instead of interacting electrons \cite{SI}.
Instead, the upward shift at $\gamma$ works in concert with weaker electron-electron repulsion matrix elements 
for states near $\gamma$ \cite{SI} that reduce their Coulomb energy penalty as the flat bands are filled.  
The ED results in this work were calculated at  twist  angle $\theta=1.1$,  interaction  strength parameter $\epsilon^{-1}=0.05$, for the $M=9$ moir\'e unit  cells system, which is sufficiently large to capture the important distinction between states near $\gamma$ and those in the rest of the 
Brillouin zone.  Unlike the model we study, experimental samples 
do exhibit clear particle-hole asymmetry.
For example, the Chern insulator states we discuss below tend to be
more prominent at positive than at negative filling factors. 
The asymmetry is thought \cite{MingHF2} to be due to nonlocal corrections to
the interlayer tunneling model we employ.  The relationship of our findings to 
experiment is addressed more fully in the discussion section below.

The many-body Hamiltonian separates into decoupled blocks 
labeled by the number of electrons in each valley $N_{K}$ and $N_{K'}$, valley-dependent
total ($S_K$ and $S_{K'}$) and azimuthal spin ($S^z_{K}$ and $S^z_{K'}$) quantum numbers, and 
total crystal momentum $(K_x,K_y)$. The separate spin quantum numbers for the two valleys apply because the model is invariant under independent valley-dependent spin rotations.

\begin{figure}[ht]
\begin{center}
\includegraphics[width=0.99\columnwidth]{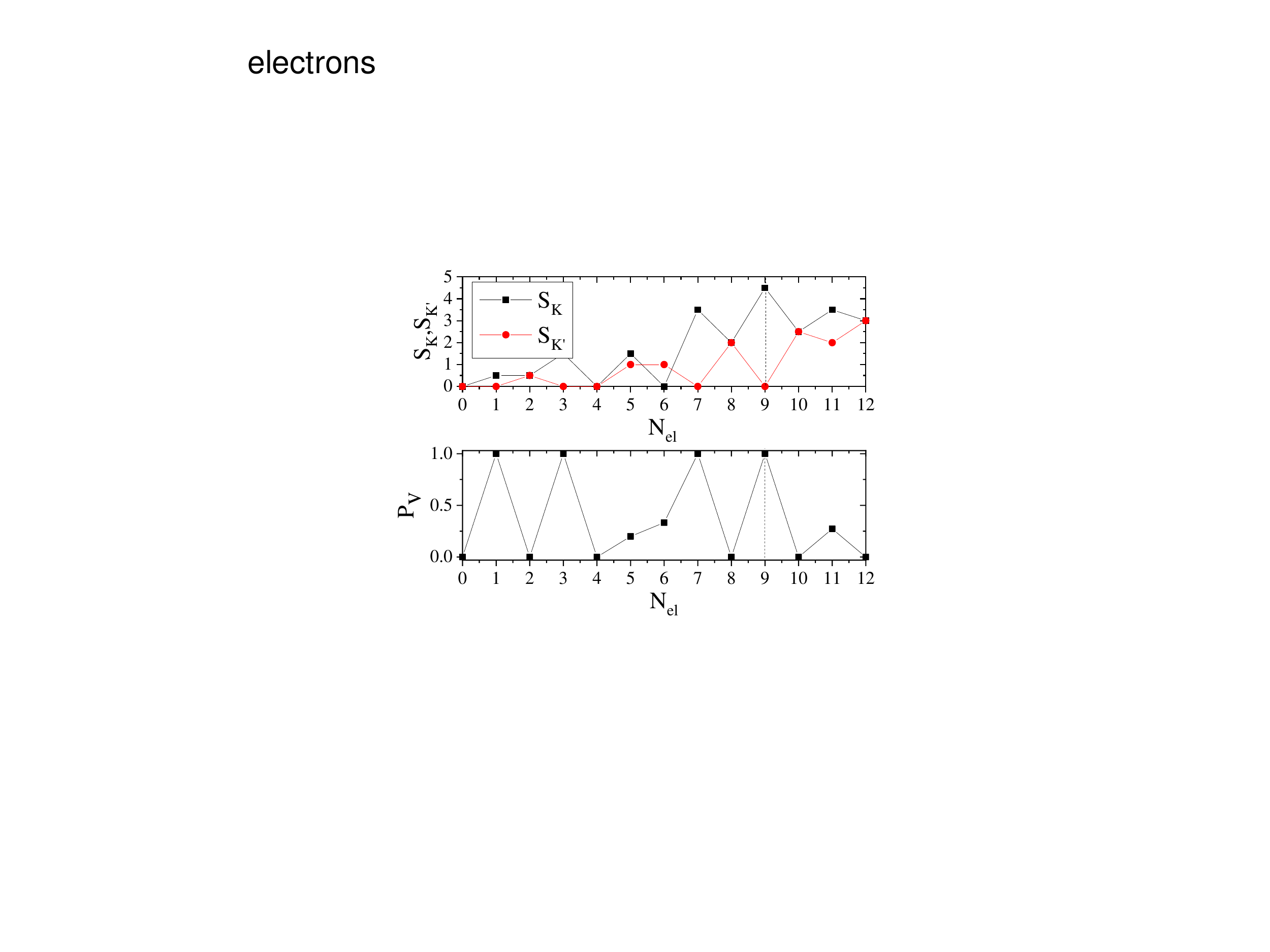}
\caption{Ground state spin and valley quantum numbers as a function of electron number 
$N_{el}$. Top: total spin $S_K$ and $S_{K'}$ in each valley. Bottom: valley polarization $P_v$. 
Integer band filling $\nu=-3$ occurs at $N_{el}=9$ highlighted by a dashed line. $S = S_K + S_{K'}$ is total spin.}
\label{Fig:PhaseDiagram_el}
\end{center}
\end{figure}

{\em Numerical results:}---
Our first important result is related to
the regime in which $|\nu| \in (2,3)$, where the ground state is commonly observed to have two 
occupied flavors.  (Our ED calculations have little access to the $|\nu|<2$ region of filling factor,
which fortunately are of lesser interest because they tend to have relatively well understood Fermi 
liquid ground state with no broken symmetries \cite{WongYazdani, ZondinerIlani,MingHF2}.)
A key issue is whether these states are fully spin polarized, or fully valley polarized, or in 
some other more complicated two-flavor state. Our ED calculations do not have access to the full Hilbert space 
across the entire $|\nu| \in (2,3)$ interval, which corresponds to the $N_{el} \in [10,18]$ in our flat-band projected ED
calculation.  For $N_{el} =10,11,12$ full Hilbert space calculations confirm that the ground state is maximally 
spin-polarized, as illustrated in Fig. \ref{Fig:PhaseDiagram_el}.  
For larger $N_{el}$ we can show that the fully spin-polarized state is lower in energy than the corresponding
fully valley-polarized state.  Some of these conclusions rest on extrapolations from calculations performed in a 
selected subspace of the full ED Hilbert space, as explained in the Supplementary Material \cite{SI}.
The conclusion that the ground state is fully spin polarized helps constrain and simplify potential 
theories of superconductivity.

For $N_{el}=9$ ($|\nu|=3$), we are able to fully explore nearly all subspaces, including all 
with particles distributed over three flavors, and subspaces with particles distributed over 
four flavors provided one of the flavors is filled by at least five particles. 
We find that the ground state is fully spin and valley polarized, and well approximated
by a single Slater determinant. For example, we find that the maximum deviation from unit momentum-state occupation
across the Brillouin zone is 0.04.
The ground state appears as a quartet with nearly degenerate doublets for each sense of
valley polarization.  By studying the response of this doublet 
to a sublattice dependent potential $m_0 \sigma_z$,
where $\sigma_z$ acts on the sublattice degree-of-freedom in both layers, 
we see that for a given valley polarization the doublet is formed by states with opposite 
sublattice polarizations and that there is observable hybridization between these
states. It is known from Hartree-Fock theory that these states are 
Chern insulators with Chern number magnitudes $|C|=1$ and signs determined by the sign of the 
product of the valley and sublattice polarization.  
In Refs. \cite{KhalafVishwanath, KhalafVishwanathZaletel} the two states with the same Chern number are 
described by a $\sigma$ model in which only the orientation of the corresponding pseudospin is retained as a relevant degree of freedom. The Chern insulator at $|\nu|=3$ \cite{BultinckZaletel, BultinckZaletel2, ZhangChern, LiuDai} 
can be viewed as a simple ferromagnet formed from these pseudospins.
From Fig.~~\ref{Fig:Fig1}(b) we conclude that $\langle \sigma_z \rangle \sim 2.25$, implying that 
$P_{sub}=\langle \sigma_z \rangle/N_{el} \sim 0.25$, in agreements with previous Hartree-Fock 
results \cite{BultinckZaletel2,MingHF}, and that 
the Hamiltonian matrix element for collective tunneling between states with opposite senses of spontaneous 
sublattice polarization (which is expected to fall exponentially with system size)
is $\sim 0.058$ meV for $N_{el}=M=9$ and (based on a separate calculation) 
$\sim 0.0094$ meV for $N_{el}=M=16$ calculation.

{\em Easy-plane valley anisotropy:}---
In Figs.~\ref{Fig:Fig1}(c) and \ref{Fig:PhaseDiagram_el} 
we see that valley polarization is completely lost when we add or remove one electron
from the $N_{el}=9$ valley and spin polarized ferromagnet.  
We attribute this behavior to the strong band splitting at $\gamma$, which has an 
outsized influence on valley anisotropy by suppressing the band-mixing degree of freedom.
An important element of our interpretation is the observation that our system has only U(1) and 
not SU(2) valley symmetry.  In the language of magnetism our system has uniaxial valley anisotropy,
which allows easy axis or easy-plane valley magnetism.  Our conclusion that the state at $\gamma$ 
plays a crucial role is supported by the property that the excitation spectra at $N=8$, where 
the $\gamma$ state is empty, and at $N=10$, where the $\gamma$ state is doubly occupied, are 
nearly identical.  Our calculations confirm that easy axis sublattice/band order is present for both
easy-axis and easy-plane valley anisotropy, with four degenerate classical states distinguished by 
the sublattice polarization of $K$ and $K'$ valley components in the easy plane case.  
The ground state responds most strongly to sublattice potentials that are identical in the two valleys,
demonstrating $AA$ or $BB$ sublattice polarizations (see SM).  These two classical states should 
have identical energies, and we conclude from the ED spectra that the tunneling between them is large at this system 
size.  We associate the excited state doublets in 
Fig.~\ref{Fig:Fig1}(c) with $AB$ and $BA$ sublattice polarizaiton for valleys $KK'$.
This interpretation is supported by strong response to valley-odd sublattice potentials.
Our ED results demonstrate that the many-particle tunneling matrix element between 
these sublattice states is greatly reduced compared to tunneling between degenerate $AA$ and $BB$ 
states.  In this case the ED spectra exhibit resonant tunneling not only between ground states, but also between
excited states of the isolated $AB$ and $BA$ sectors.   

\begin{figure}
\begin{center}
\includegraphics[width=0.99\columnwidth]{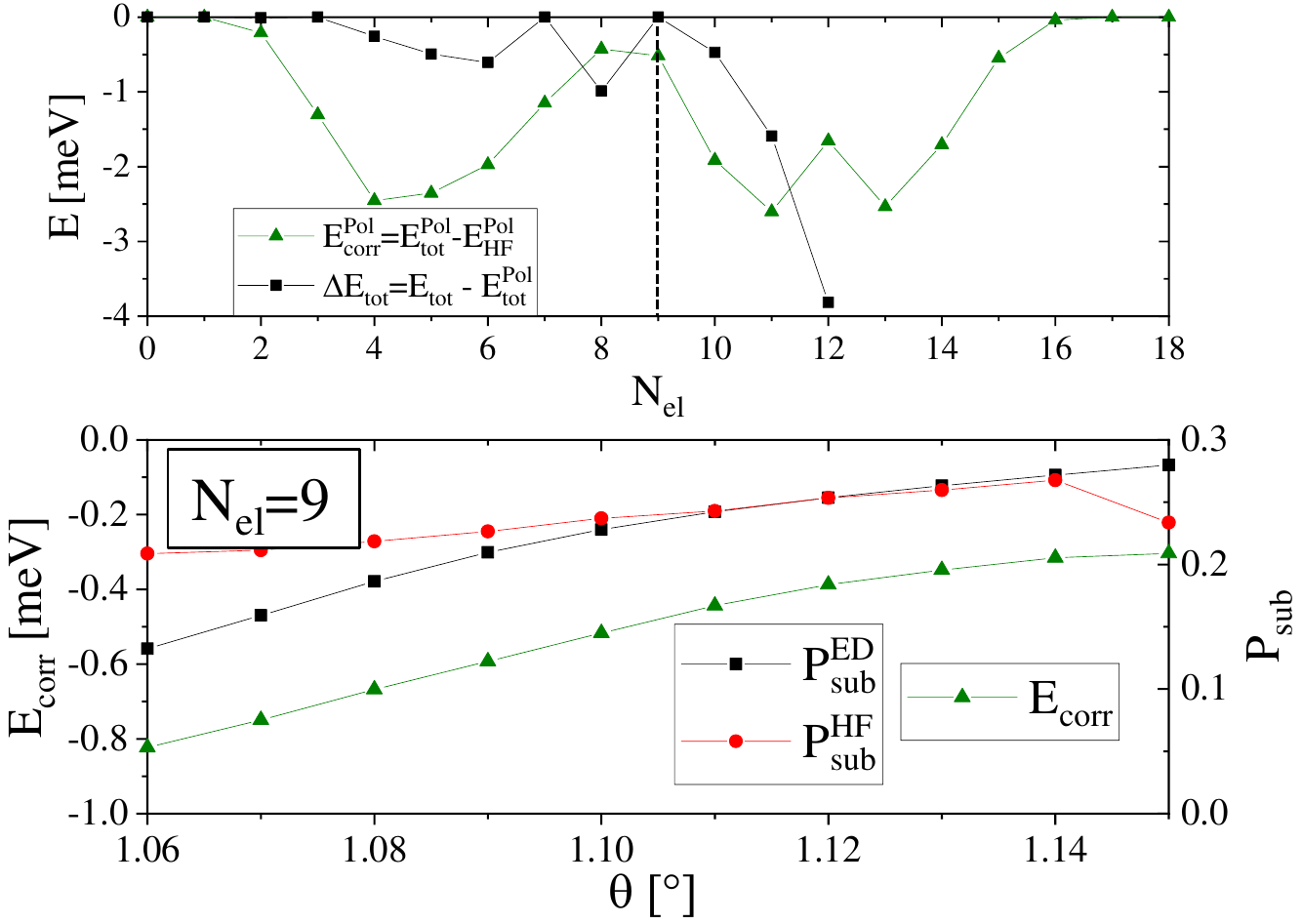}
\caption{The correlation energy per moir\'e period as a function of filling factor (top) 
and as a function of the twist angle for $N_{el}=9$ together with corresponding sublattice polarizations (bottom). Within this range of angles, both exact diagonalization and Hartree-Fock calculations predict full flavor polarization. Top: the green triangles correspond to one flavor full-spin-polarization calculations and $E^{Pol}_{corr}= E^{Pol}_{tot} -  E^{Pol}_{HF}$, where  $E^{Pol}_{tot}$ is the ground state energy from exact diagonalization calculations and $E^{Pol}_{HF}$ is the mean-field energy from self-consistent Hartree-Fock calculations. The black squares show to the energy difference between the ground state energy and the lowest energy
state in the full flavor polarization sector.  The black dashed line marks the filling factor $\nu=-3$. Bottom: the twist angle dependence at $N_{el}=9$ with green triangles for the correlation energy and 
black squares and red circles for ED and HF sublattice polarizations.}
\label{Fig:Corr_ener}
\end{center}
\end{figure}

{\em Discussion:}---
Our calculations show that MATBG ground state energies are generally speaking well approximated by 
unrestricted Hartree-Fock approximations that allow spin, valley, and sublattice symmetries to be broken.
In the top panel of Fig.~\ref{Fig:Corr_ener} we show the dependence of the correlation energy on electron
number in the subspace with full spin and valley flavor polarization over the full range of available filling factor
for that flavor between $\nu_f=-1$ and $\nu_f=1$.
The correlation energy, defined as the difference between the ED ground state energy and the minimum 
energy single-Slater determinant, vanishes when the orbital doublet is empty ($\nu_f=-1$) and full ($\nu_f=1$), and 
also reaches an extremely small value in the insulating state at $\nu_f=0$.  
These results suggest insulating states at all integer filling factors with a band filling per flavor equal to 
-1, 0, or 1 are accurately rendered by Hartree-Fock calculations, even when symmetry is broken by choosing 
different band filling factors for different flavors. 
For a given total integer filling factor a variety of different states,
characterized by different flavor-dependent filling factors and senses of sublattice polarization,
are expected to compete closely in energy.  The states have different total Chern numbers
with $|C|=1$ for $|\nu|=3$, 0 or 2 for $|\nu|=2$, 1 or 3 for $|\nu|=1$, and 0, 2 or 4 for $\nu=0$. (The $|\nu|=1$ and $|\nu|=0$ cases are outside of the reach of ED.) 
In Fig.~\ref{Fig:Corr_ener} we see that the correlation energy is larger away from integer filling factors.
We expect that this trend will be stronger in sectors with less flavor polarization, and that 
Hartree-Fock calculations therefore overestimates the tendency to break flavor symmetries.
Insulating states at $|\nu|=1$, will therefore compete with metallic states with no broken flavor 
symmetries that have much larger correlation energies.  
The difference in energy between the true ground state and the ground state in the fully polarized 
sector increases quickly for $N_{el}>9$, showing that the energy cost of valley polarization
quickly increases.

The appearance of insulating states at integer filling factors depends on 
screening environment, twist angle, and band structure details that we have not fully explored here.
For example in the bottom panel of Fig.~\ref{Fig:Corr_ener} we illustrate how
the correlation energy of the $\nu=-3$ Chern insulator state depends on twist angle.
As expected the correlation energy is reduced as the twist angle increases 
relative to the magic angle.  Surprisingly though, the sublattice polarization
increases and is more accurately estimated by Hartree-Fock calculations as twist angle increases \cite{MingHF}.
Evidently the physics responsible for the broken symmetries is basically that of 
exchange interactions, which are captured by Hartree-Fock calculations, with correlations working against order. While finite size effects are present in all ED calculations, a direct comparison with Hartree-Fock results for the same system size (Fig. \ref{Fig:Corr_ener}) and results extrapolated
to the thermodynamic limit shown in the Supplemental Material confirm this conclusion.
Like the twist angle $\theta$, the screening parameter $\epsilon$ used in our calculations influences quantitative conclusions.
Interactions in the flat bands of MATBG are screened by the surrounding hexagonal boron nitride (hBN)
dielectric, by the nearby electrical gates, and by transitions between the flat 
and remote bands \cite{PizarroWehling}.  
Strictly speaking, the latter two effects yield wave-vector-dependent contributions
to the dielectric constant with gates dominating in importance at small wavevectors and screening within the 
bilayer dominating a larger wave vectors. 

{\em Acknowledgment:} 
A. H. M. was supported by the U.S. Department of 339 Energy, Office of Science, Basic Energy Sciences, under 340 Award No. DE-SC0022106. 
P.P. acknowledges financial support by the Polish National Agency for Academic Exchange (NAWA). The authors acknowledge helpful
interactions with A. Bernevig, N. Regnault, T. Senthil, A. Vishwanath, and F. Wu. ED calculations were performed 
using the Wrocław Center for Networking and Supercomputing and the Texas Advanced Computing Center (TACC).

\newpage
\appendix


\begin{thebibliography}{99}

\bibitem{MorellFlatBand} 
E. Suarez Morell, J. D. Correa, P. Vargas, M. Pacheco and Z. Barticevic, 
Phys. Rev. B \textbf{82}, 121407(R) (2010).

\bibitem{BMModel} 
R. Bistritzer and A. H. MacDonald,
Proc. Natl. Acad. Sci. U.S.A. \textbf{108}, 12233 (2011).

\bibitem{KharitonovMaxim}
Maxim Kharitonov,
Phys. Rev. B \textbf{86}, 195435 (2012).  

\bibitem{ZhangNiuMacDonald}
F. Zhang, J. Jung, G. A. Fiete, Q. Niu, and A. H. MacDonald,
Phys. Rev. Lett. \textbf{106}, 156801 (2011).

\bibitem{MacDonaldJungZhang}
A. H. MacDonald1, J. Jung, and F. Zhang,
Phys. Scr. \textbf{2012}, 014012 (2012).

\bibitem{BarlasMacDonald}
Y. Barlas, R. Cote, K. Nomura, A. H. MacDonald,
Phys. Rev. Lett. \textbf{101}, 097601 (2008).

\bibitem{Tutuc2017} 
K. Kim, A. DaSilva, S. Huang, B. Fallahazad, S. Larentis, T. Taniguchi, K. Watanabe, B. J. LeRoy, A. H. MacDonald, and E. Tutuc,
Proc. Natl. Acad. Sci. U.S.A. \textbf{114}, 3364 (2017).

\bibitem{CaoInsulator} 
Y. Cao, V. Fatemi, A. Demir, S. Fang, S. L. Tomarken, J. Y. Luo, 
J. D. Sanchez-Yamagishi, K. Watanabe, T. Taniguchi, E. Kaxiras, R. C. Ashoori, and P. Jarillo-Herrero, 
Nature (London) \textbf{556}, 80 (2018).


\bibitem{CaoSuper} 
Y. Cao, V. Fatemi, S. Fang, K. Watanabe, T. Taniguchi, E. Kaxiras, and P. Jarillo-Herrero, 
Nature (London) \textbf{556}, 43 (2018).

\bibitem{YoungDean} 
M. Yankowitz, S. Chen, H. Polshyn, K. Watanabe, T. Taniguchi, D. Graf, A. F. Young, and C. R. Dean, 
Science \textbf{363}, 11059 (2019).

\bibitem{MacDonaldEfetov} 
X. Lu, P. Stepanov, W. Yang, M. Xie, M. Ali Aamir, I. Das, C. Urgell,
K. Watanabe, T. Taniguchi, G. Zhang, A. Bachtold, A. H. MacDonald and D. K. Efetov,
Nature (London) \textbf{574}, 656 (2019).

\bibitem{UriJarilloHerrero} 
A. Uri, S. Grover, Y. Cao, J. A. Crosse, K. Bagani, D. Rodan-Legrain, Y. Myasoedov, K. Watanabe, T. Taniguchi, P. Moon, M. Koshino, P. Jarillo-Herrero, and E. Zeldov,
Nature (London) \textbf{581}, 47 (2020).

\bibitem{BalentsYoung}
L. Balents, C. R. Dean, D. K. Efetov and A. F. Young,   
Nat. Phys. \textbf{16}, 725–733 (2020).

\bibitem{WongYazdani}
D. Wong, K. P. Nuckolls, M. Oh, B. Lian, Y. Xie, S. Jeon, K. Watanabe, T. Taniguchi, B. A. Bernevig and A. Yazdani,
Nature (London) \textbf{582}, 198–202 (2020).

\bibitem{ZondinerIlani}
U. Zondiner, A. Rozen, D. Rodan-Legrain, Y. Cao, R. Queiroz, T. Taniguchi, K. Watanabe, Y. Oreg, F. von Oppen, Ady Stern, E. Berg, P. Jarillo-Herrero and S. Ilani
Nature (London) \textbf{582}, 203–208 (2020).


\bibitem{SharpeGoldhaber}
A. L. Sharpe, E. J. Fox, A. W. Barnard, J. Finney, K. Watanabe, T. Taniguchi, M. A. Kastner, and D. Goldhaber-Gordon,
Science \textbf{365}, 605 (2019).

\bibitem{SerlinYoung}
M. Serlin, C. L. Tschirhart, H. Polshyn, Y. Zhang, J. Zhu, K. Watanabe, T. Taniguchi, L. Balents, and A. F. Young,
Science \textbf{367}, 900 (2020).

\bibitem{StepanovEfetov}
P. Stepanov, I. Das, X. Lu, A. Fahimniya, K. Watanabe, T. Taniguchi, F. H. L. Koppens, J. Lischner, L. Levitov, and D. K. Efetov,
Nature (London) \textbf{583}, 375 (2020).

\bibitem{WuAndrei}
S. Wu, Z. Zhang, K. Watanabe, T. Taniguchi, and E. Y. Andrei,
Nat. Mater. \textbf{20}, 488 (2021).

\bibitem{DasEfetov}
I. Das, X. Lu, J. Herzog-Arbeitman, Z.-D. Song, K. Watanabe, T. Taniguchi, B. A. Bernevig, and D. K. Efetov,
Nat. Phys. \textbf{17}, 710 (2021).

\bibitem{ParkFlavor}
J.  M.  Park, Y.  Cao, K. Watanabe, T. Taniguchi, and P.  Jarillo-Herrero
Nature (London) \textbf{592}, 43 (2021).

\bibitem{AroraNadjiPerge}
H. S. Arora, R. Polski, Y. Zhang, A. Thomson, Y. Choi, H. Kim, Z. Lin, I. Z.Wilson, X. Xu, J.-H. Chu, K.Watanabe, T. Taniguchi, J. Alicea, and S. Nadj-Perge, 
Nature (London) \textbf{583}, 379 (2020).

\bibitem{CaoNematicity}
Y. Cao, D. Rodan-Legrain, J. Min Park, F. N. Yuan, K. Watanabe, T. Taniguchi, R. M. Fernandes, L. Fu, and P. Jarillo-Herrero,
Science \textbf{372}, 264 (2021).

\bibitem{StepanovCompeting}
P. Stepanov, M. Xie, T. Taniguchi, K. Watanabe, X. Lu, A. H. MacDonald, B. Andrei Bernevig, and D. K. Efetov,
arXiv:2012.15126

\bibitem{RozenIlani}
A. Rozen, J. M. Park, U. Zondiner, Y. Cao, D. Rodan-Legrain, T. Taniguchi, K. Watanabe, Y. Oreg, Ady Stern, E. Berg, P. Jarillo-Herrero and S. Ilani
Nature (London) \textbf{592}, 214 (2021).

\bibitem{FuModel} 
N. F. Q. Yuan and L. Fu,
Phys. Rev. B \textbf{98}, 045103 (2018).


\bibitem{VishwanathOrigin} 
H. C. Po, L. Zou, A. Vishwanath, and T. Senthil,
Phys. Rev. X \textbf{8}, 031089 (2018).

\bibitem{Scalettar} 
H. Guo, X. Zhu, S. Feng, and R. T. Scalettar,
Phys. Rev. B \textbf{97}, 235453 (2018). 

\bibitem{Phillips} 
B. Padhi, C. Setty, and P. W. Phillips,
Nano Lett. \textbf{18}, 6175 (2018).

\bibitem{SpinLiquid} 
V. Y. Irkhin and Y. N. Skryabin,
JETP Lett. \textbf{107}, 651 (2018).

\bibitem{WangModel} 
J. F. Dodaro, S. A. Kivelson, Y. Schattner, X.-Q. Sun, and C. Wang,
Phys. Rev. B \textbf{98}, 075154 (2018).

\bibitem{MottAF} 
T. Huang, L. Zhang, and T. Ma,
Sci. Bull. \textbf{64(5)}, 310 (2019).

\bibitem{Yang} 
C.-C. Liu, L.-D. Zhang, W.-Q. Chen, and F. Yang,
Phys. Rev. Lett. \textbf{121}, 217001 (2018)

\bibitem{PALee} 
X. Y. Xu, K. T. Law, and P. A. Lee,
Phys. Rev. B \textbf{98}, 121406(R) (2018).

\bibitem{Vafek} 
J. Kang and O. Vafek,
Phys. Rev. X \textbf{8}, 031088 (2018).

\bibitem{Rademaker} 
L. Rademaker and P. Mellado,
Phys. Rev. B \textbf{98}, 235158 (2018).

\bibitem{KoshinoFu} 
M. Koshino, N. F. Q. Yuan, T. Koretsune, M. Ochi, K. Kuroki, and L. Fu,
Phys. Rev. X \textbf{8}, 031087 (2018).

\bibitem{Bascones} 
J. M. Pizarro, M. J. Calderon, and E. Bascones,
J. Phys. Commun. \textbf{3}, 035024 (2019).

\bibitem{Ochi} 
M. Ochi, M. Koshino, and K. Kuroki,
Phys. Rev. B \textbf{98}, 081102(R) (2018).

\bibitem{FuJune} 
H. Isobe, N. F. Q. Yuan, and L. Fu,
Phys. Rev. X \textbf{8}, 041041 (2018).

\bibitem{AFMonHoneyComb} 
A. Thomson, S. Chatterjee, S. Sachdev, and M. S. Scheurer,
Phys. Rev. B \textbf{98}, 075109 (2018).

\bibitem{GuineaInteractions} 
F. Guinea and N. R. Walet, 
Proc. Natl. Acad. Sci. U.S.A. \textbf{115}, 13174 (2018).

\bibitem{MingHF} 
Ming Xie, Allan H. MacDonald,
Phys.  Rev.  Lett. \textbf{124}, 097601 (2020).

\bibitem{MingHF2} 
Ming Xie, Allan H. MacDonald,
arXiv:2010.07928v1.

\bibitem{BultinckZaletel} 
N. Bultinck, S. Chatterjee, and M. P. Zaletel,
Phys. Rev. Lett. \textbf{124}, 166601 (2020).

\bibitem{BultinckZaletel2} 
N. Bultinck, E. Khalaf, S. Liu, S. Chatterjee, A. Vishwanath,and M. P. Zaletel,
Phys. Rev. X \textbf{10}, 031034 (2020).

\bibitem{KhalafVishwanath} 
E. Khalaf, S. Chatterjee, N. Bultinck, M. P. Zaletel, and A. Vishwanath, 
Sci. Adv.  \textbf{7}, 19 (2021).

\bibitem{KhalafVishwanathZaletel} 
E. Khalaf, N. Bultinck, A. Vishwanath, and M. P. Zaletel,
arXiv:2009.14827.

\bibitem{ChatterjeeZaletel} 
S. Chatterjee, N. Bultinck, and M. P. Zaletel,
Phys. Rev. B \textbf{101}, 165141 (2020).

\bibitem{ChatterjeeZaletel2} 
S. Chatterjee, M. Ippoliti, and M. P. Zaletel,
arXiv:2010.01144 

\bibitem{ParkerZaletelBultinck} 
D. E. Parker, T. Soejima, J. Hauschild, M. P. Zaletel, and Nick Bultinck,
Phys. Rev. Lett. \textbf{127}, 027601 (2021).

\bibitem{XuBalents} 
C. Xu and L. Balents,
Phys. Rev. Lett. \textbf{121}, 087001 (2018).

\bibitem{FuSuper} 
F. Wu, A. H. MacDonald, and I. Martin,
Phys. Rev. Lett. \textbf{121}, 257001 (2018).

\bibitem{Juricic} 
B. Roy and V. Juricic,
Phys. Rev. B \textbf{99}, 121407(R) (2019). 

\bibitem{Das} 
S. Ray, J. Jung, and T. Das,
Phys. Rev. B \textbf{99}, 134515 (2019).

\bibitem{Super1} 
T. J. Peltonen, R. Ojajarvi, and T. T. Heikkila,
Phys. Rev. B \textbf{98}, 220504(R) (2018).

\bibitem{Super2} 
Y.-Z. You and A. Vishwanath,
Quantum Mater. \textbf{4}, 16 (2019).

\bibitem{Super3} 
X.-C. Wu, , K. A. Pawlak, C.-M. Jian, and C. Xu,
arXiv:1805.06906.

\bibitem{Super4} 
M. Fidrysiak, M. Zegrodnik, and J. Spalek,
Phys. Rev. B \textbf{98}, 085436 (2018).

\bibitem{SantosEarly} 
J. M. B. Lopes dos Santos, N. M. R. Peres, and A. H. Castro Neto,
Phys. Rev. Lett. \textbf{99}, 256802 (2007).

\bibitem{VishwanathWannierObstructions} 
H. C. Po, H. Watanabe, and A. Vishwanath,
Phys. Rev. Lett. \textbf{121}, 126402 (2018).

\bibitem{VishwanathFaithful} 
H. C. Po, L. Zou, T. Senthil, and A. Vishwanath,
Phys. Rev. B \textbf{99}, 195455 (2019).

\bibitem{TarnopolskyVishwanath} 
G. Tarnopolsky, A. J. Kruchkov, and A. Vishwanath, 
Phys. Rev. Lett. \textbf{122}, 106405 (2019).

\bibitem{LiuDai} 
J. Liu, J. Liu, and X. Dai,
Phys. Rev. B \textbf{99}, 155415 (2019).

\bibitem{ZhangChern}
Y.-H. Zhang, D. Mao, Y. Cao, P. Jarillo-Herrero, and T. Senthil, 
Phys. Rev. B \textbf{99}, 075127 (2019).

\bibitem{Semenoff}
G. W. Semenoff, 
Phys. Scr. \textbf{T146}, 014016 (2012).

\bibitem{Jung2014} 
J. Jung, A. Raoux, Z. Qiao, and A. H. MacDonald, 
Phys. Rev. B \textbf{89}, 205414 (2014).

\bibitem{SaitoYoung} 
Y. Saito, J. Ge, K. Watanabe, T. Taniguchi, and A. F. Young,
Nat. Phys. \textbf{16}, 926 (2020).

\bibitem{AjeshCollectiveModel}
A. Kumar, M.  Xie, and  A.  H. MacDonald
Phys. Rev. B \textbf{104}, 035119 (2021).



\bibitem{TBGI}
B. A. Bernevig, Z. D. Song, N. Regnault, and B. Lian,
Phys. Rev. B \textbf{103}, 205411 (2021).

\bibitem{TBGII}
Z.-D. Song, B. Lian, N. Regnault, and B. A. Bernevig,
Phys. Rev. B \textbf{103}, 205412 (2021).

\bibitem{TBGIII}
B. A. Bernevig, Z. D. Song, N. Regnault, and B. Lian,
Phys. Rev. B \textbf{103}, 205413 (2021).

\bibitem{TBGIV}
B. Lian, Z. D. Song, N. Regnault, D. K. Efetov, A. Yazdani, and B. A. Bernevig, 
Phys. Rev. B \textbf{103}, 205414 (2021).

\bibitem{TBGV}
B. A. Bernevig, B. Lian, A. Cowsik, F. Xie, N. Regnault, and Z. D. Song,
Phys. Rev. B \textbf{103}, 205415 (2021).

\bibitem{TBGVI}
F. Xie, A. Cowsik, Z. D. Song, B. Lian, B. A. Bernevig, and N. Regnault, 
Phys. Rev. B \textbf{103}, 205416 (2021).

\bibitem{KangVafek}
Oskar Vafek, and Jian Kang,
Phys. Rev. Lett. \textbf{125}, 257602 (2020).

\bibitem{PizarroWehling}
J. M. Pizarro, M. Rosner, R. Thomale, R. Valenti, and T. O. Wehling,
Phys. Rev. B \textbf{100}, 161102(R) (2019).

\bibitem{KangVafek2}
Jian Kang, and Oskar Vafek 
Phys. Rev. Lett. \textbf{122}, 246401 (2019).

\bibitem{RepellinSenthil} 
C. Repellin, Z. Dong, Y.-H. Zhang, and T. Senthil 
Phys. Rev. Lett. \textbf{124}, 187601 (2020).

\bibitem{LiaoQMC}
Y. Da Liao, Z. Y. Meng, and X. Y. Xu
Phys. Rev. Lett. \textbf{123}, 157601 (2019).

\bibitem{SoejimaParkerZaletel} 
T. Soejima, D. E. Parker, N. Bultinck, J. Hauschild, and M. P. Zaletel 
Phys. Rev. B \textbf{102}, 205111 (2020).

\bibitem{KangVafek3}
J. Kang and O. Vafek, 
Phys. Rev. B \textbf{102}, 035161 (2020).




\bibitem{SI}
See the Supplemental Materials for additional filling factor dependent results.

\end{thebibliography}
\end{document}


\title{Supplementary materials \\ Exact Diagonalization for Magic Angle Twisted Bilayer Graphene}
\author{Pawel Potasz}
\affiliation{Department of Physics, The University of Texas at Austin, Austin, TX 78712, USA}
\affiliation{Department of Physics, Wroclaw University of Science and Technology, 50-370 Wroclaw, Poland} 
\author{Ming Xie}
\author{A. H. MacDonald}
\affiliation{Department of Physics, The University of Texas at Austin, Austin, TX 78712, USA}

\maketitle

\section{TBG Quasiparticle energy bands }
\begin{figure}[ht]
\begin{center}
\includegraphics[width=0.99\columnwidth]{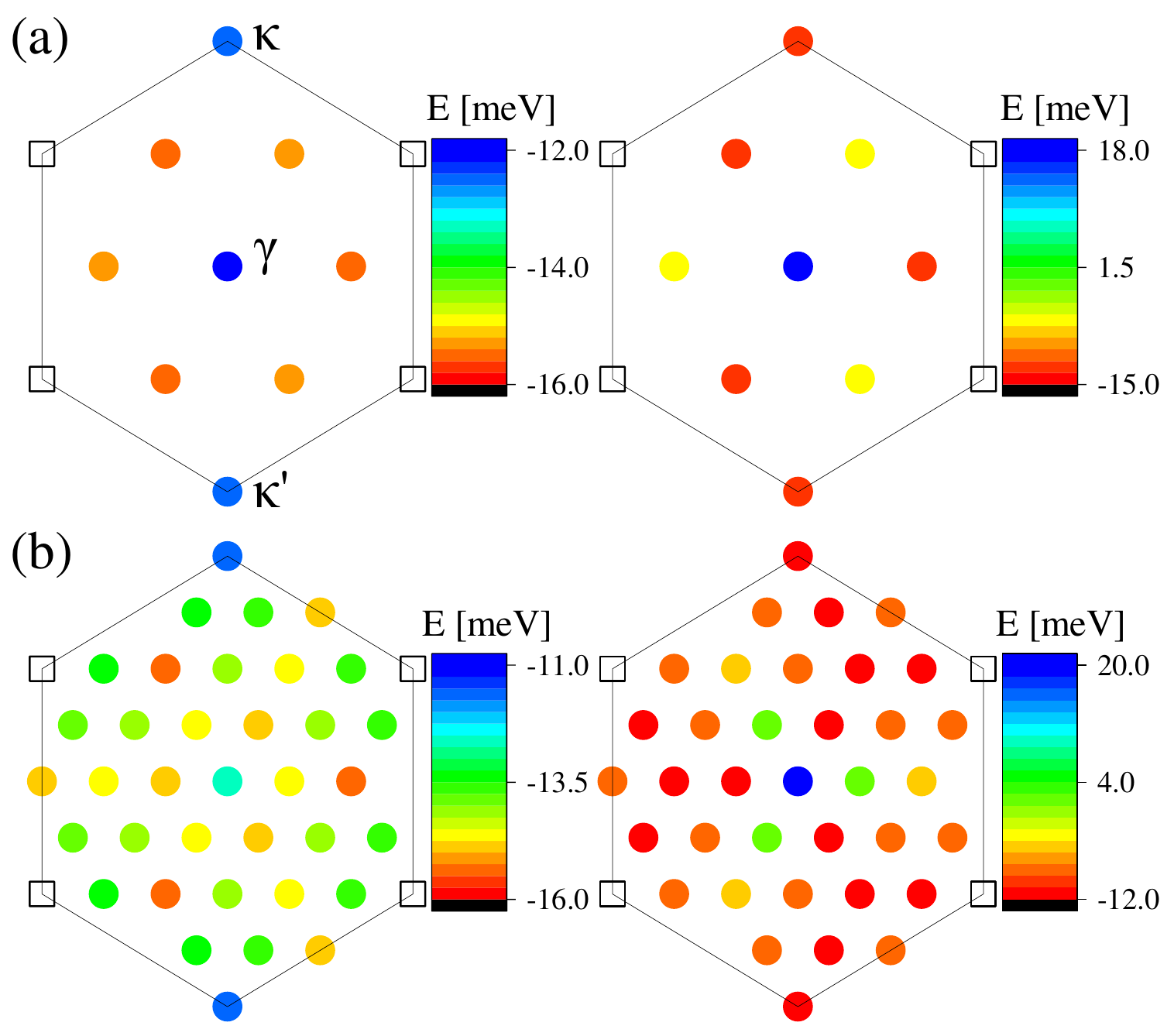}
\caption{Quasiparticle energies of empty flat bands for 
(a) $M=9$ ($N_x = 3$, $N_y = 3$) and (b) 
$M=36$ ($N_x = 6$, $N_y = 6$) moir\'e unit cell systems at twist angle $\theta=1.1$,
interaction strength parameter $\epsilon^{-1}=0.05$, and inter-layer intrasublattice 
and intersublattice hopping parameters $\omega_1=110$ meV and $\omega_0=0.8\omega_1$.
The valence band is on the left and the conduction band on the right. Empty squares are used to mark equivalent momenta.}
\label{Fig:FiniteSizeBands}
\end{center}
\end{figure}

Fig.~\ref{Fig:FiniteSizeBands} plots the empty-system valence and conduction bands
obtained by diagonalizing $\epsilon_{i} \delta_{i',i}+ \Sigma_{i',i}$ 
for finite size systems containing (a) $M=9$ and (b) $M=36$ moir\'e unit cells at twist angle $\theta=1.1$,
interaction strength parameter $\varepsilon^{-1}=0.05$, and inter-layer hopping parameters 
$\omega_1=110$ meV amd $\omega_0=0.8\omega_1$ for inter-sublattice and  
intrasublattice tunneling.  (Each band has a two-fold spin-degeneracy and two valley partners ($K$ $K'$)
related by time-reversal symmetry (see Ref. [\onlinecite{MingHF}] for a more detailed 
discussion of band and screening ($\epsilon$) parameters). 
We see in Fig.~\ref{Fig:FiniteSizeBands} that the main features of the flat bands are captured
already at $M=9$, used for most of our ED calculations.  The valence band is extremely flat with a width less than 
$5$ meV and a local maximum at the moir\'e Brillouin-zone $\gamma$-point, 
The conduction band is much broader with a width larger than $25$ meV and 
a sharp peak at $\gamma$.  Importantly the energy separation between conduction and valence 
bands at $\gamma$ is comparable to the largest energy scales in the problem.
The conduction band width greatly exceeds the few meV width \cite{KoshinoFu} obtained 
when the remote band self-energy is neglected, but does not strongly influence the band separation. 
The bands have a three-fold rotational symmetry which is more clearly revealed
for $M=36$. 

\section{Coulomb matrix elements}
\begin{figure}
\begin{center}
\includegraphics[width=0.99\columnwidth]{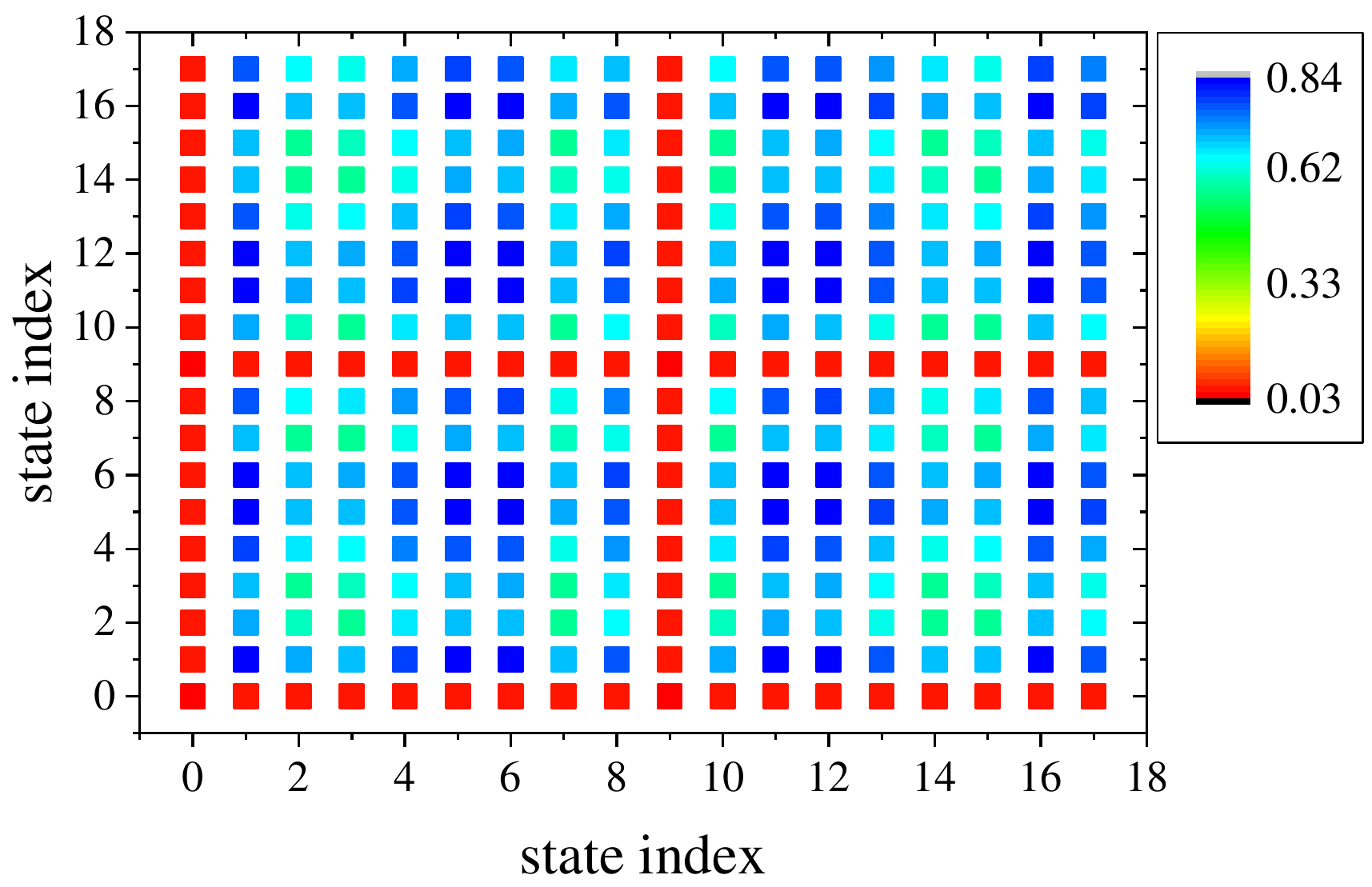}
\includegraphics[width=0.99\columnwidth]{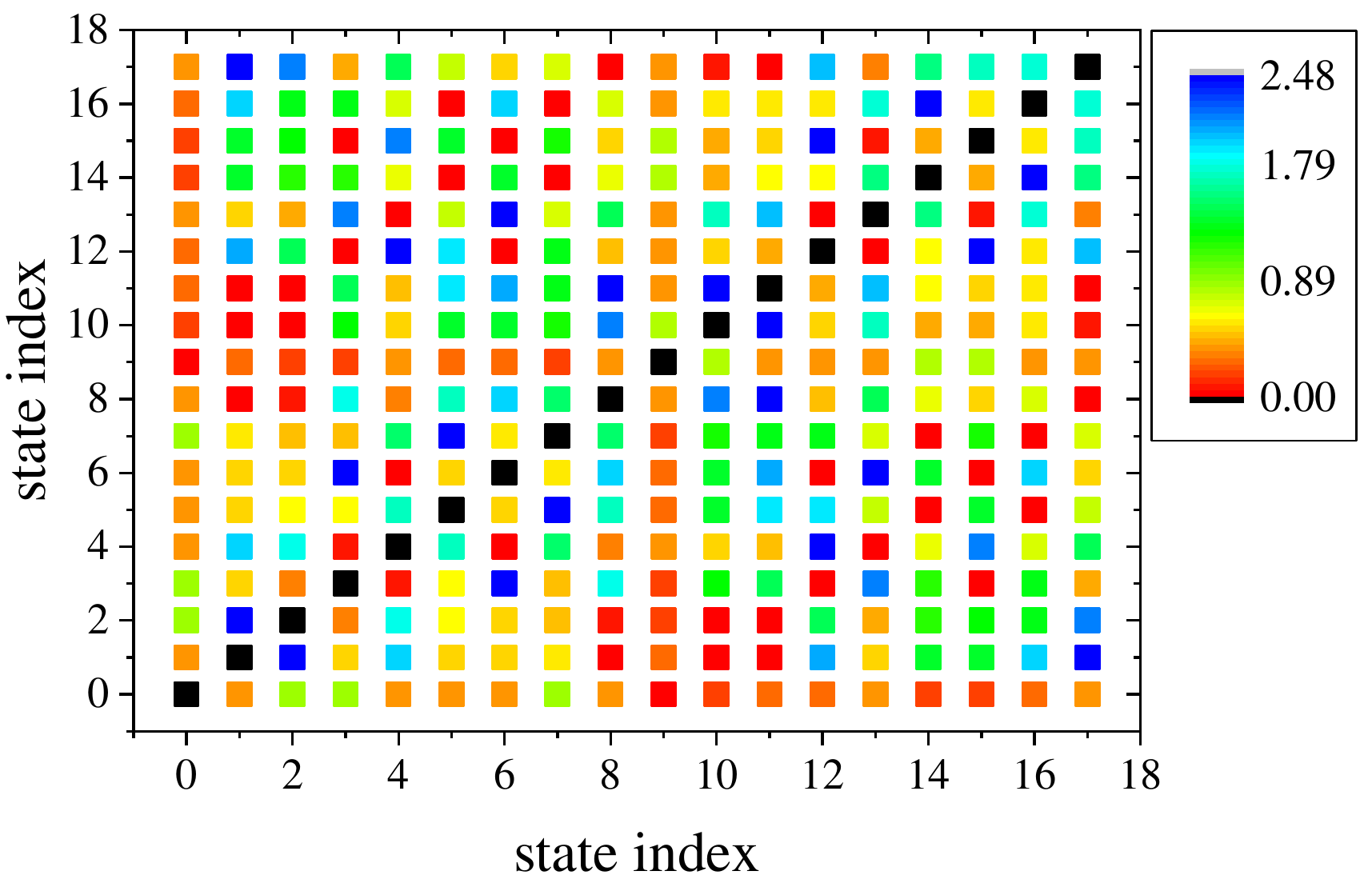}
\caption{A map of direct Coulomb (top) and exchange (bottom) matrix elements within the flat bands. State indices $\langle0,8\rangle$ correspond to the valence band with a state index $i$, $i = k_yN_{x}+k_x$, and $\langle 9,17\rangle$ to the conduction band with a state index $i$, $i = N_{x}*N_{y} + k_yN_{x}+k_x$. The $\gamma$ points correspond to the state index $i=0$ for the valence band and $i=9$ for the conduction band. Direct Coulomb matrix elements from $i=0$ and $i=9$ to all the states have clearly lower values in comparison to all other matrix elements.}
\label{Fig:Coul_el}
\end{center}
\end{figure}

Exact diagonalization (ED) calculations account for all direct and exchange terms, 
and also for the many valley and momentum conserving processes in which pairs of 
electrons are scattered between different occupation number states.  
In Fig. \ref{Fig:Coul_el} direct (top) and exchange (bottom) Coulomb matrix elements 
are shown as a function of their single-particle state momentum and orbital labels.
We label the single-particle states by $i = k_yN_{x}+k_x$ for the valence band and 
$i = N_{x}*N_{y}+ k_yN_{x}+k_x$ for the conduction band, where
${\bf{k}} =\frac{k_x}{N_x}{\bf b}_{1}+\frac{k_y}{N_y}{\bf b}_{2}$, ${\bf b}_{1,2}$ are reciprocal lattice vectors, 
and $N_x$ and $N_y$ are the integers that specify the numbers of unit cells in the finite size 
system in the primitive lattice vector directions ($M=N_x\times N_y$ is a total number of unit cells).
Here we use the terms valence and conduction bands to refer respectively to the lower and 
higher energy states obtained by diagonalizaing the single-particle Hamiltonian including the 
remote band self-energy contribution.
The state indices $i=0$ and $i=9$ correspond to the $\gamma$ point.
Clearly direct interactions with these states are weaker than all other Coulomb matrix elements. 
This reflects the fact that wavefuncions around the $\gamma$ point are more spread in a real space comparing to $k$-points from other regions of the Brillouin zone. In the map of exchange interactions one can distinguish stronger 
interaction between states with momentum ${\bf k}$ and $-{\bf k}$ indicated by a blue color with values above $2.4$ meV. Interactions within the second valley, and between valleys,  are obtained by performing time-reversal operations 
on these figures.

\section{Sublattice polarizations and band occupations}
A bilayer has four sublattices $\alpha = \{ A_1, B_1,A_2,B_2\}$, where $A,B$ distinguish 
the two sublattices of one graphene honeycomb layer and $1,2$ is a layer index.
The sublattice density is defined as
 \begin{eqnarray}
n_\alpha ({\bf r}) = \langle\Psi_{\alpha}^{+}({\bf r})\Psi_{\alpha} ({\bf r})\rangle,
\label{Chargedistr}
\end{eqnarray}
where the field operators are
 \begin{eqnarray}
\Psi^{+}_\alpha({\bf r})= \frac{1}{\sqrt{M}}\sum_{{\bf k},n}  \phi^{*}_{{\bf k},n,\alpha}({\bf r}) a^{+}_{{\bf k},n,\alpha},
\label{field}
\end{eqnarray}
and $a^+_{{\bf k},n,\alpha}$ are creation operators of particles with momentum ${\bf k}$ and band index $n$, and $\phi^{*}_{{\bf k},n}({\bf r})$ are Bloch wavefunctions 
expressed in a plane wave basis of moir\'e reciprocal lattice vectors ${\bf G}$:
 \begin{eqnarray}
\phi_{{\bf k}, n,\alpha}({\bf r})= \sum_{{\bf G}}  z^n_{\alpha,{\bf G}}({\bf k}) e^{i({\bf k}+{\bf G}){\bf r}},
\label{Psik_four}
\end{eqnarray}
where $z^{n}_{\alpha,{\bf G}}({\bf k})$ are the expansion coefficients of the 
continuum model spinors.  

The many-body state is defined 
as $|\psi^{MB}_l\rangle=\sum_s A^{l}_s |s\rangle$ with coefficients $A^{l}_s$ obtained from diagonalization Hamiltonian matrix, where $|s\rangle=\prod_{i_s} a^{+}_{{\bf k}_{i_s}}|0\rangle$
are occupation number configurations corresponding to Slater determinant (SD)s wavefunctions. 
Using the definitions
above, the density per moir\'e cell on sublattice $\alpha$ is
\begin{equation}
n^{MB}_\alpha = 
 \frac{1}{M} \sum_{n_s,n_p} \sum_{{\bf k,G}}  z^{n_s}_{{\bf G},\alpha}({\bf k}) z^{n_p}_{{\bf G},\alpha}({\bf k}) \;  A^{*l}_s A^l_p (-1)^{r} 
\label{sublat_occup}
\end{equation}
where $z^{n_s}$ are coefficients of an occupied band $n_s$ of $s$-th basis configuration, and a factor $(-1)^r$ comes from anticommutation of operators and $r$ is a number of exchanges for a given set $(n_s,n_p)$. 
The sublattice polarization is defined as
 \begin{eqnarray}
P_{sub} = \frac{|n^{MB}_{A_1}+n^{MB}_{A_2}-n^{MB}_{B_1}- n^{MB}_{B_2}|}{\sum_{\alpha} n^{MB}_{\alpha}}.
\label{Psub}
\end{eqnarray}
We can also define a total valence band occupation $n_{VB}$, by 
restricting indices $n_s$ to ones from the valence band only in Eq. \ref{sublat_occup} 
and summing over all sublattices $\alpha$. With these definitions, $P_{sub}$ and $n_{VB}$ are quantities per particle, thus normalized to one in the case of full sublattice polarization or full valence band occupation.

All sublattices are equivalent and an external field is necessary to 
break the symmetry between them. In order to induce a finite sublattice polarization, 
a sublattice mass term is added, described by the Hamiltonian
 \begin{eqnarray}
\text{H}_{m_0} = \sum_{{\bf k},n,\alpha,\mu} m^{\mu}_0 \; 
\sigma^{\alpha\alpha '}_z a^{+}_{{\bf k},n,\alpha } a_{{\bf k},n,\alpha ' },
\label{Hmass}
\end{eqnarray}
where momentum ${\bf k}$ is restricted to a single valley $\mu=K,K'$, $\sigma_z$ is 
a Pauli matrix in the $z$ direction in space of sublattices and we consider two cases, 
one with same mass $m_0=m^{K}_0=m^{K'}_0$ and one with 
opposite  mass terms $m_0=m^{K}_0=-m^{K'}_0$ in the two valleys. 

\begin{figure}
\begin{center}
\includegraphics[width=0.99\columnwidth]{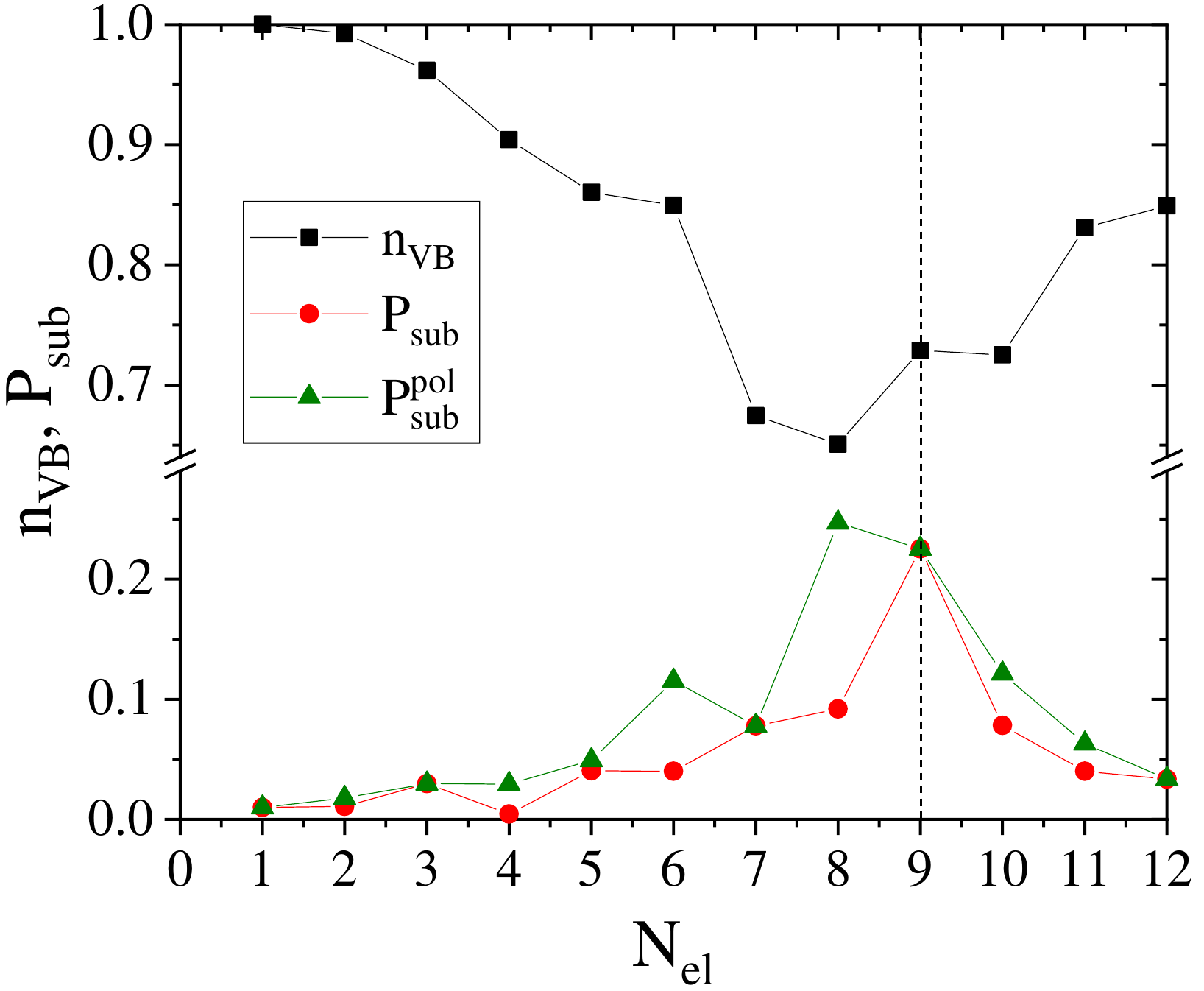}
\caption{The valence band occupation $n_{VB}$ and sublattice polarization $P_{sub}$ of the many-body ground state obtain  from exact diagonalization as a function of the filling factor for a sublattice mass $m_0=1.0$ meV. For a comparison, sublattice polarization $P^{pol}_{sub}$ for a flavor polarized lowest energy state is also shown. A vertical dash line indicates an integer filling factor $\nu=-3$ corresponding to $N_{el}=9$ electrons.}
\label{Fig:Occup}
\end{center}
\end{figure}

Fig. \ref{Fig:Occup} shows the total valence band occupation $n_{VB}$ (black squares) and 
the sublattice polarization of the ground state as a function of a particle number $N_{el}$.
(Here the valence band is defined as the lower band of flat band states when the remote band 
self-energy is included in the single-particle Hamiltonian.) 
We evaluate sublattice polarization $\text{P}_{\text{sub}}$ for
the ED many-body ground state (red circles) and for the case in which only one flavor is occupied
(green triangles), both calculated for a finite sublattice mass $m_0=1$ meV. 
For a small filling, the mixing between the valence and conduction bands is weak.  
For $N_{el}<7$ the fractional valence band occupation is more than 80$\%$. 
For these fillings there is no clear flavor polarization suggesting electron gas like behavior. 
This is also seen in sublattice polarization $\text{P}_{\text{sub}}$, which is close to zero, confirming 
approximately equal filling of all aublattices, and four flavors. 
Moving closer to integer filling $\nu=-3$ for $N_{el}=9$, the valence band occupation
fraction drops below $n_{VB}=0.7$ already at $N_{el}=7$, with a tiny increase of sublattice polarization, and spin polarization (see MS).  Interactions start playing the central role as the integer 
filling factor is approached.  At $N_{el}=9$ the ground state has clear 
spontaneous sublattice polarization
as emphasized in the main MS.  This is seen in these relatively large $m_0$ calculations 
as a peak in $\text{P}_{\text{sub}}$. Similar behavior is observed for the
single flavor case (green triangles), except at $N_{el}=8$, where the peak in the 
sublattice polarization is the highest. This is related to a formation of a 
state with a uniform filling of all momentum points except the one at $\gamma$, which remains nearly empty.
This behavior is a reminder that the interactions of particles filling in $\gamma$ with other electrons are 
significantly weaker compared to interactions between particles filling other momenta.
Further studies will be needed to determine how universal these details are when the 
twist angle and the band structure model are changed, however we anticipate that the details of 
sublattice and valley order are sensitive to changes in band filling factor that change occupation
numbers near $\gamma$, where both the wavefunctions and the dispersion are anomalous.  
Above integer filling $\nu=-3$, sublattice polarizations deceases, and the 
fractional valence band occupation increases.

\section{Filling $\nu=-3$}
\begin{figure}
\begin{center}
\includegraphics[width=0.99\columnwidth]{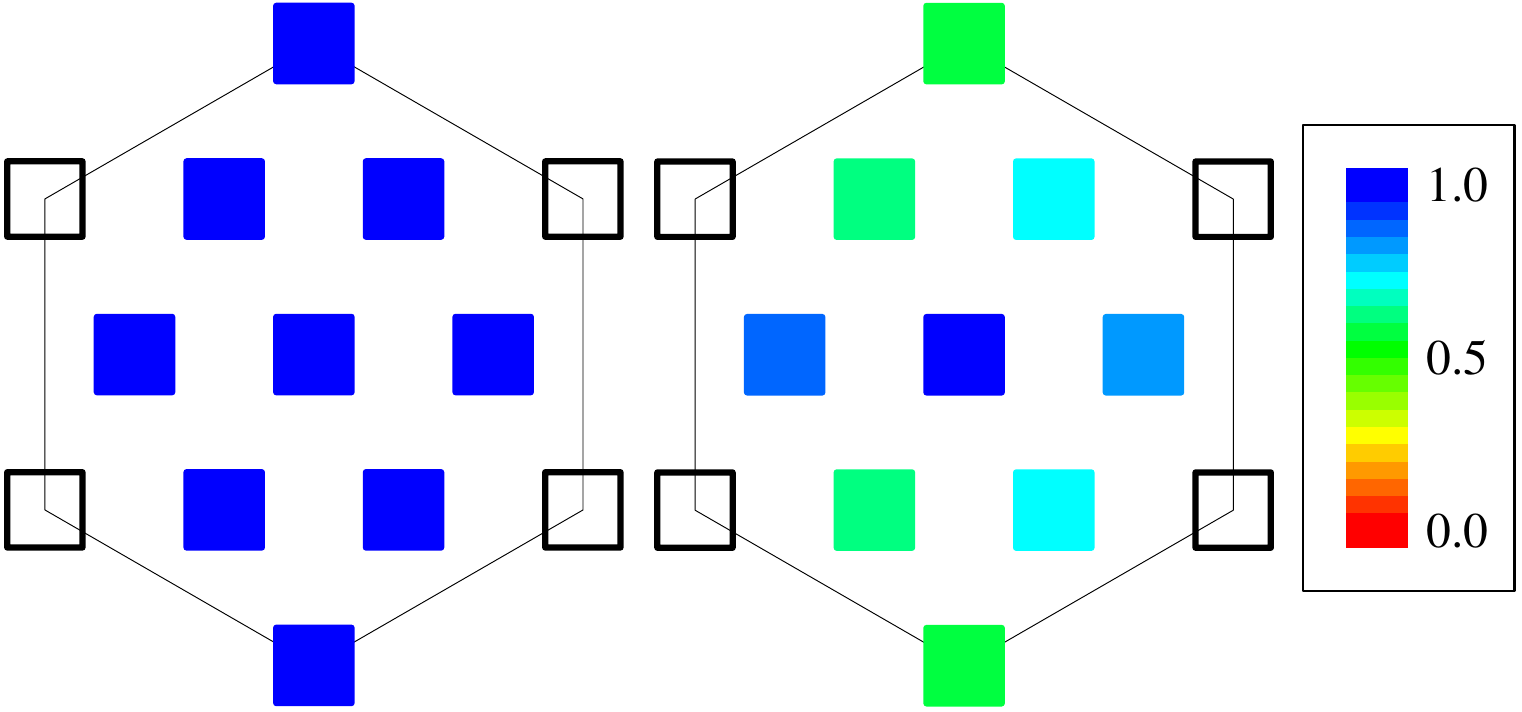}
\caption{Total flat band occupation (left) and valence band occupation (right) for the $N_{el}=9$ ground state. The filled squares correspond to the discrete ${\bf k}$-points of the 
$M=9$ finite-size system and empty squares to the corresponding points in neighboring Brillouin zones.}
\label{Fig:9el}
\end{center}
\end{figure}

The insulating state for $N_{el}=9$ is a spontaneous sublattice symmetry broken state. In Fig. \ref{Fig:9el} we show the total flat band occupation (left) and the occupation 
projected onto valence band states (right). 
The deviation from equal occupation of each momentum is less than 0.04.  
However the valence band occupation changes strongly with momentum, indicating that the ground 
state consists of k-dependent nontrivial mixtures of the original valence and conduction band states. 

\begin{figure}
\begin{center}
\includegraphics[width=0.99\columnwidth]{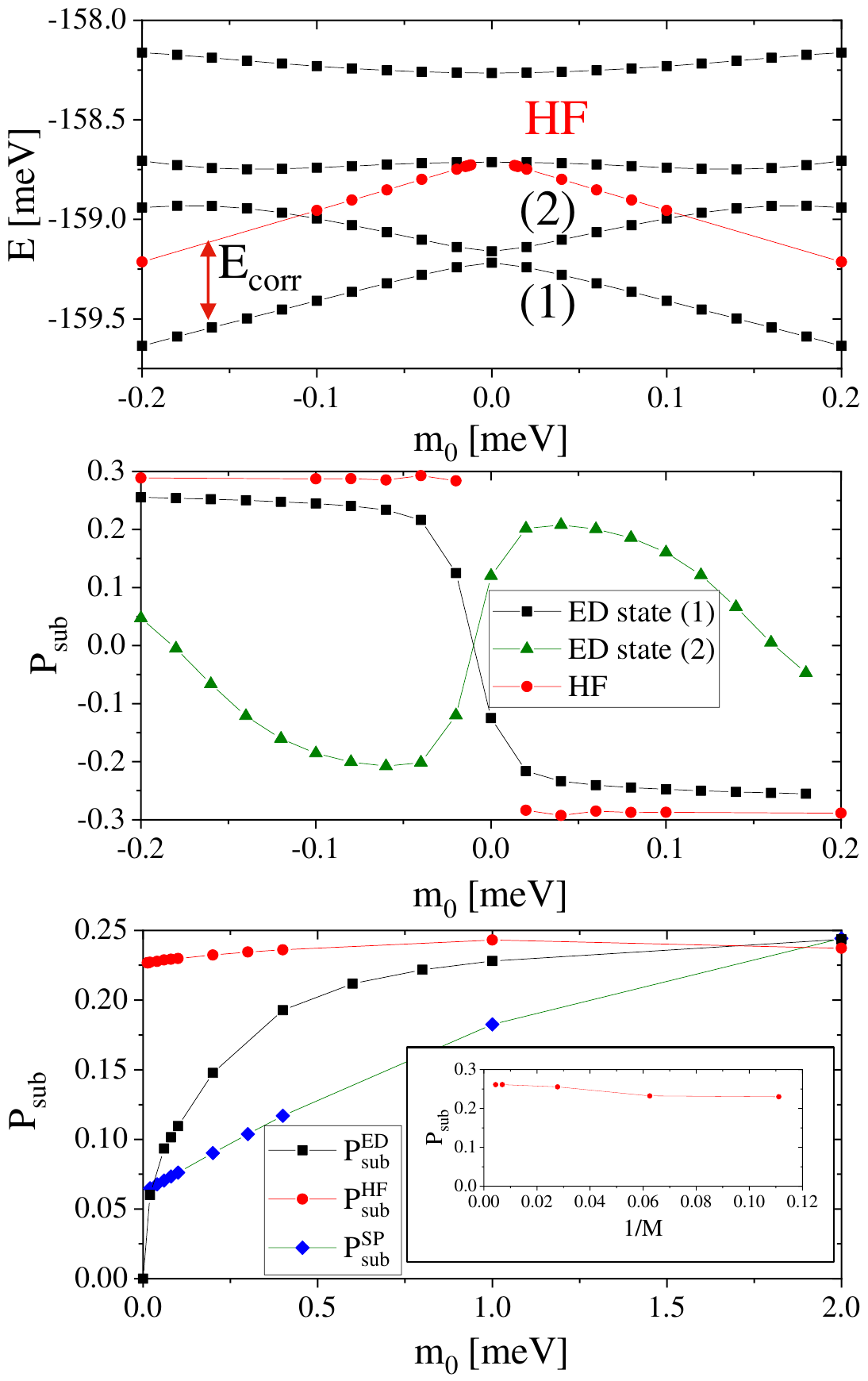}
\caption{Top panel: Total energy as a function of a the sublattice mass term for
four low-energy many-body states obtained from exact diagonalization (ED) (black squares) and Hartree-Fock (HF) energy (red circles). The vertical arrow shows the correlation energy $E_{corr}$. 
Middle panel: Sublattice polarization calculated using the relation $P_{sub}=(\frac{dE}{dm_0})/N_{el}$.  Bottom panel: Sublattice polarization $P_{sub}$ as a function of the sublattice mass term for ED, HF, and single particle (SP) ground states calculated using the definition $P_{sub}=\langle\sigma_z\rangle$. The inset show extrapolation of HF sublattice polarization to a thermodynamic limit.}
\label{Fig:Psub}
\end{center}
\end{figure}

We analyze the response of the four lowest many-body energy states to a sublattice potential mass term in the
top panel of Fig. \ref{Fig:Psub}. The ground state is nearly double degenerate,  with a splitting related to collective 
tunneling between states with opposite senses of spontaneous 
sublattice polarization that should vanish in a thermodynamic limit. These two lowest energy 
states reveal linear dependence of energy on mass, 
while two upper states remain relatively unaffected by the sublattice field, as also shown in the main MS. 
We show also the corresponding HF energy dependence on $m_0$, obtained for the same $M=9$ system size.
The comparison allows us to determine the correlation energy $E_{corr}$, 
indicated by a vertical red arrow, which remains almost constant with $m_0$. The magnitude of the 
sublattice polarization can be determined by $P_{sub}=(\frac{dE}{dm_0})/N_{el}$ (middle panel). The curve corresponding to the lowest state (black squares) quickly saturates at $P^{ED}_{sub}\sim 0.25$, while the curve for the second state (green triangles) reaches $P_{sub}\sim 0.2$ for $m_0=0.02$ and drops down with further increases 
in $m_0$ due to level-repulsion by the third state. 
The HF results have slightly larger $P^{HF}_{sub}\sim 0.3$ 
and are almost sublattice mass independent.
One can compare the results above with sublattice polarization obtained from the definition $P_{sub}=\langle\sigma_z\rangle$ (bottom panel). Sublattice polarizations from ED smoothly increases with $m_0$, approaching the HF result for sufficiently large $m_0$. 
The blue diamonds illustrate the single particle (SP) sublattice polarization. In summary, from both definitions of sublattice polarization, and both methods, ED and HF, sublattice polarization is $P_{sub}\sim 0.25$, and close to a value in a thermodynamic limit obtained from HF (inset in the bottom panel). 

\begin{figure}
\begin{center}
\includegraphics[width=0.99\columnwidth]{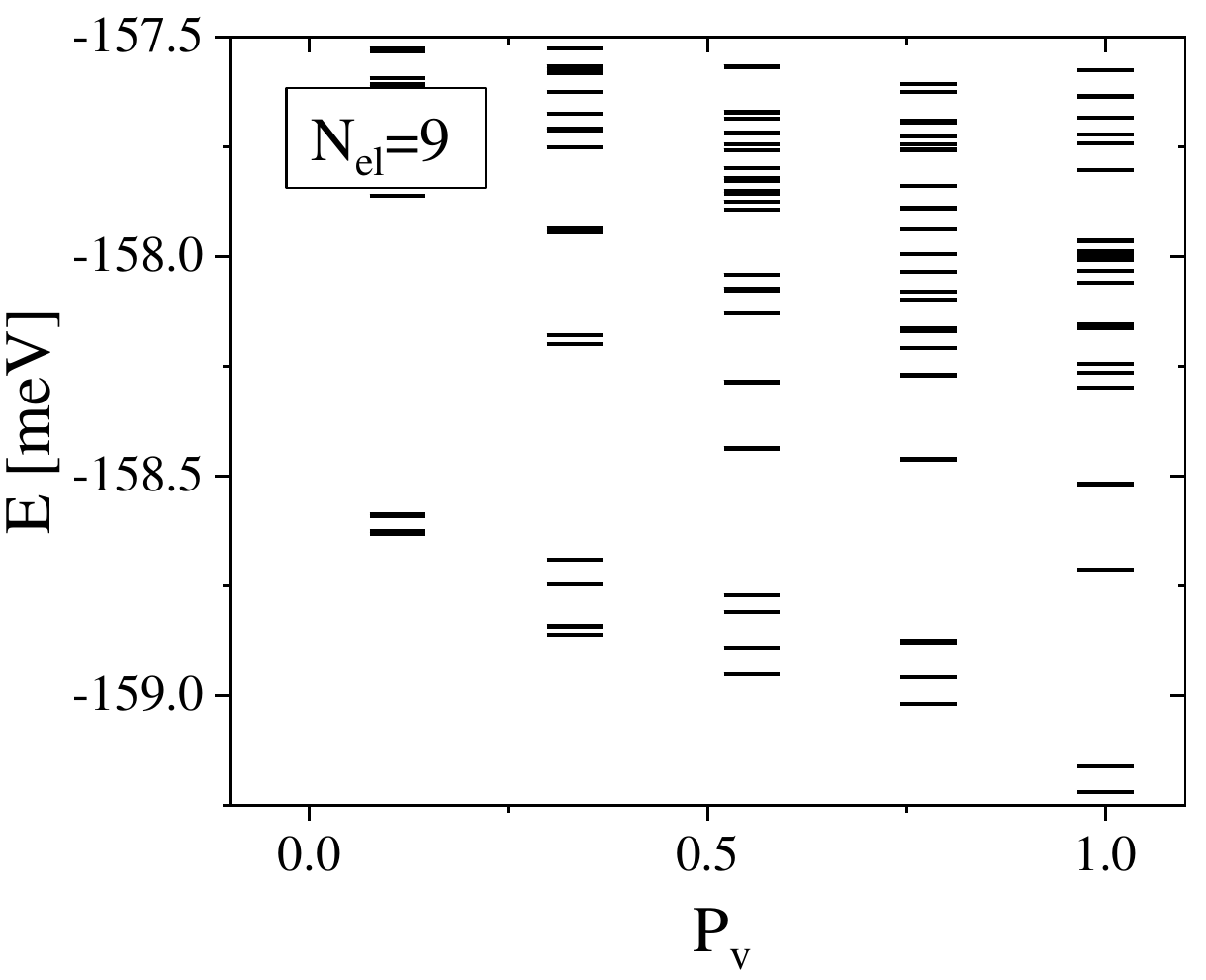}
\caption{Low energy many-body spectrum as a function of valley polarization $P_v$ for $N_{el}=9$.}
\label{Fig:9spec}
\end{center}
\end{figure}
Fig. \ref{Fig:9spec} shows the low energy many-body spectrum for $N_{el}=9$ as a function of valley polarization $P_{v}$. The ground state is maximally flavor polarized, and energy gradually increases with a decrease of valley polarization. 
The energy gap seen between the four lowest energy states and higher energy states is an artifact of plotting only the 
four lowest states from every total momentum subspace.  The energy gap for $P_{v}=1$ between the lowest energetic doublet, below $E=-159.0$ meV and the third state around $E=-158.7$ meV is correctly rendered.
Sets of pairs of nearly degenerate states are seen in different valley polarization sectors, but responses to mass fields
are needed to identify their character more completely.  We notice that e.g. two lowest states for $P_v=0.33$ do not exhibit a linear response to mass fields, while the energy states around $E=158.7$ do, and for both, even and odd, mass fields, which is related to unequal distribution of particles over two valleys ($N_K=6$, $N_{K'}=3$).  

\section{One particle charge excitation from filling $\nu=-3$}
\begin{figure}
\begin{center}
\includegraphics[width=0.99\columnwidth]{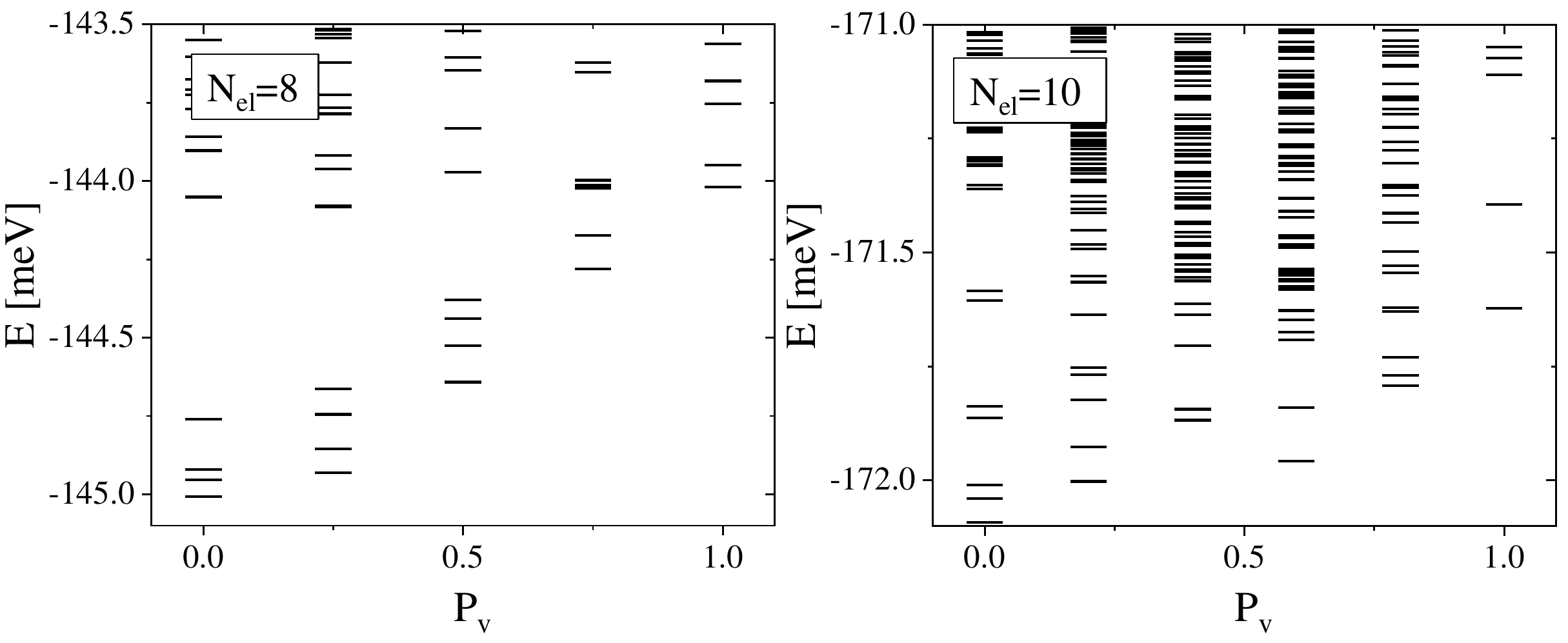}
\caption{Low energy many-body spectrum as a function of valley polarization $P_v$ for $N_{el}=8$ (left) and $N_{el}=10$ (right).}
\label{Fig:810spec}
\end{center}
\end{figure}

\begin{figure}
\begin{center}
\includegraphics[width=0.99\columnwidth]{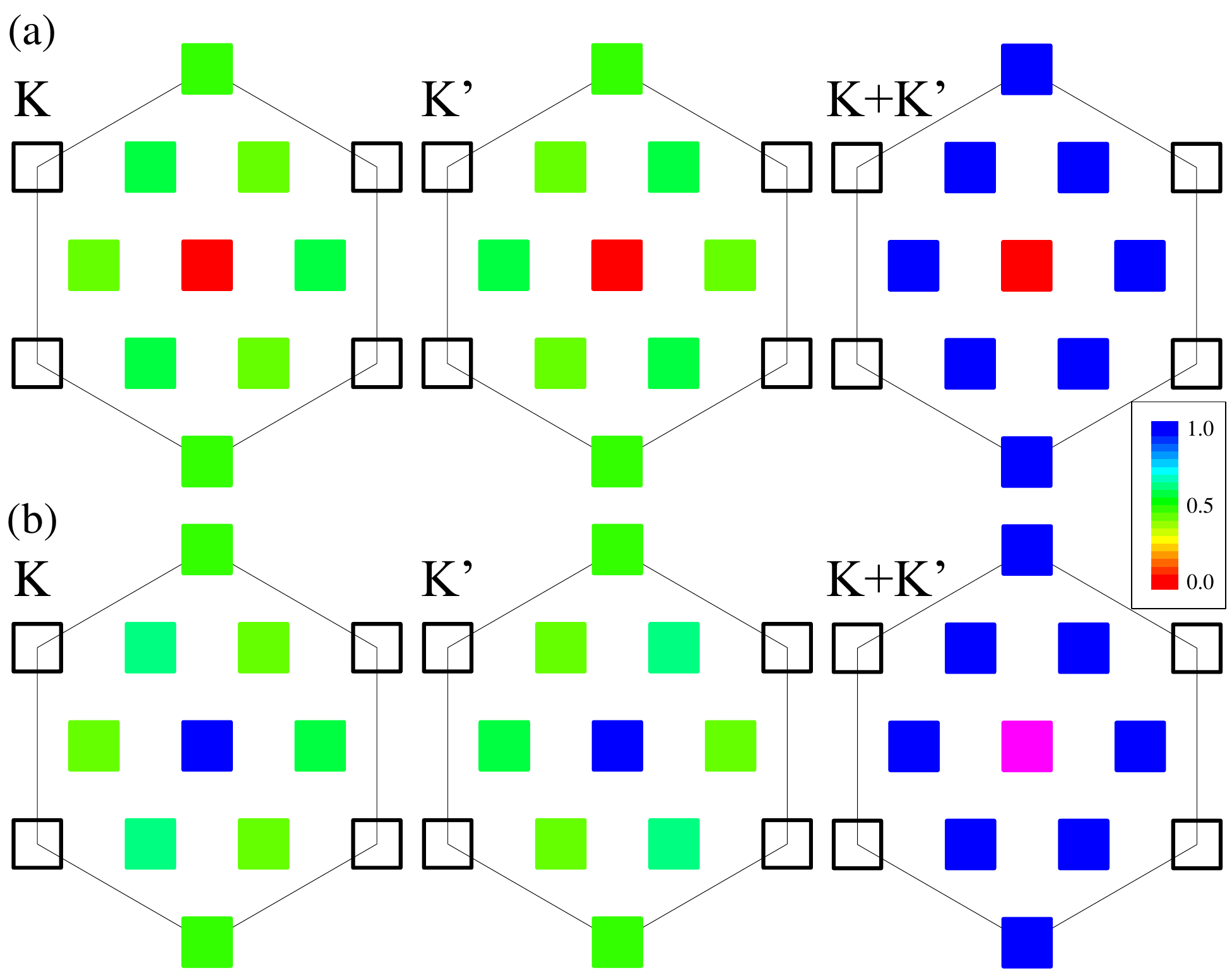}
\caption{Total flat band occupation  of the ground state for (a) $N_{el}=8$ and (b) $N_{el}=10$, for valley $K$ (left), valley $K'$ (middle), and both valleys $K+K'$ (right). An occupation indicated by a purple color for $N_{el}=10$ is close to two, $\sim 1.94$. Filled squares represent $M=9$ $k$-points from a discretized first Brillouin zone and empty squares to corresponding points from neighboring Brillouin zones.}
\label{Fig:810occup}
\end{center}
\end{figure}

\begin{figure}
\begin{center}
\includegraphics[width=0.99\columnwidth]{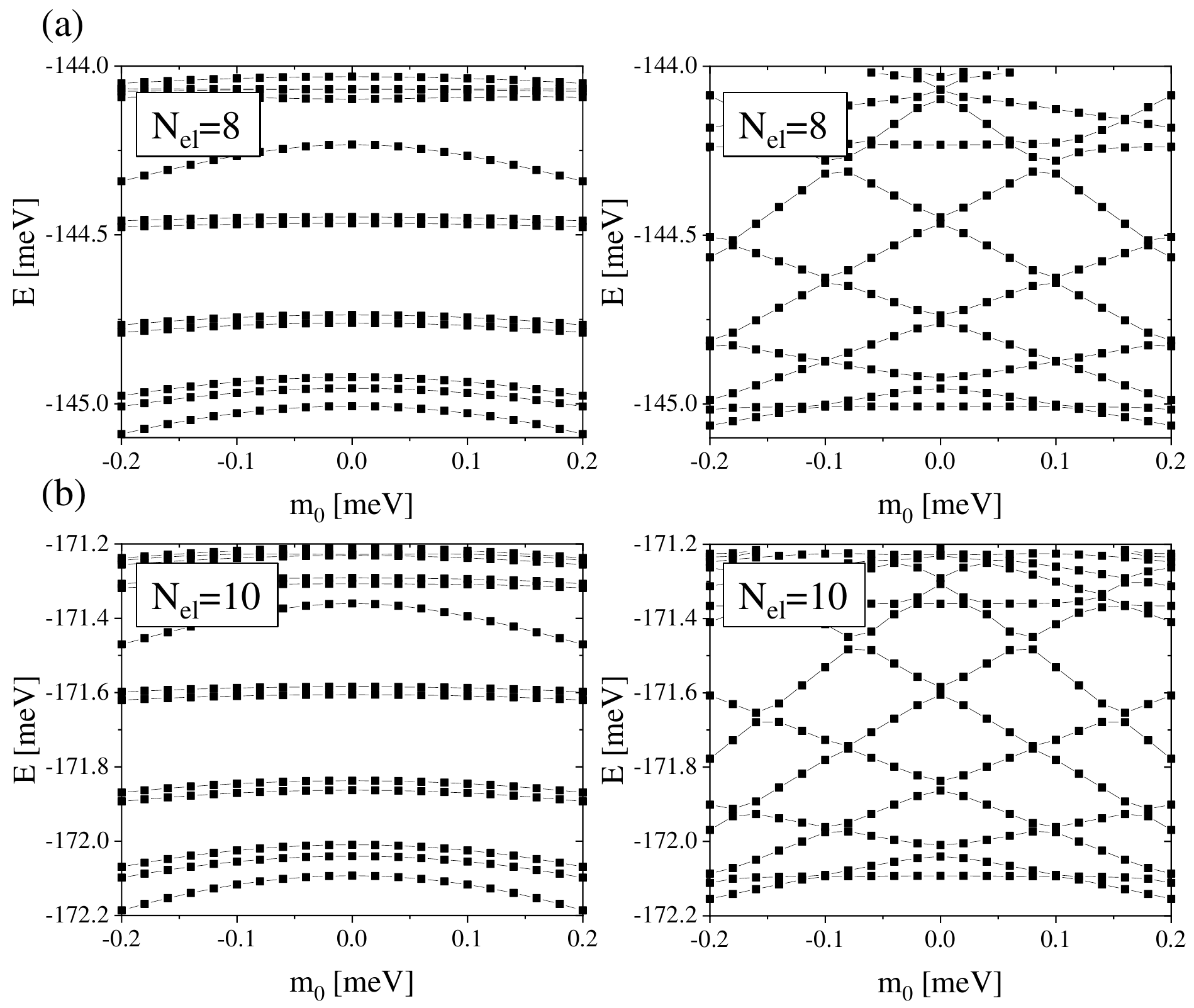}
\caption{A sublattice mass dependence of the (a) $N_{el}=8$ and (b) $N_{el}=10$ four lowest energy states with the same mass field $m_0=m^{K}_0=m^{K'}_0$ (left) and opposite mass field $m_0=m^{K}_0=-m^{K'}_0$ (right).}
\label{Fig:Fig_810mass}
\end{center}
\end{figure}

We investigate charge excitation from the $\nu=-3$ state by removing, $N_{el}=8$, or adding, $N_{el}=10$, one particle. In Fig. \ref{Fig:810spec} many-body spectra as a function of valley polarization are shown. Both types of charge excitation have fully spin polarized and fully valley unpolarized ground states. For $N_{el}=8$, the energy of lowest excitations gradually increases with valley polarization, and a quite similar spectrum is seen for $N_{el}=10$, with the largest energetic cost of the lowest excitation for full valley polarization. Similarity between $\pm 1$ particle excitation are also seen in corresponding ground state quasi particle  momentum occupations, shown in Fig. \ref{Fig:810occup}. For $N_{el}=10$, the two extra particles added to the system with $N_{el}=8$, fill $\gamma$ points from the two valleys, and occupations of other momenta remain almost unchanged. This is consistent with the Coulomb matrix elements, where the particles filling the $\gamma$ point interact very weakly 
with particles at other momenta. Total occupation summed over two valleys is close to one for $\pm 1$ particle excitation for nonzero momenta, deviating from one by less than $\sim 0.001$ for $N_{el}=8$ and $\sim 0.011$ for $N_{el}=10$. The occupation indicated by the purple color for $N_{el}=10$ is close to two, $\sim 1.94$. 
In Fig. \ref{Fig:Fig_810mass} low energy sublattice mass dependence for (a) $N_{el}=8$  and (b) $N_{el}=10$ for the same (left) and opposite (right) sublattice mass fields in two valleys are shown. The lowest state slightly response to the even mass field and is completely not affected by the odd mass field. This is an evidence of the character of this state, with equal occupation of the opposite sublattice in two valleys, and slight imbalance between two sublattice within a valley. Three doublets of higher energy states show a clear pattern in the odd mass field, a linear response, and energy separation between them that increases approximately quadratically, and the splitting within a doublet decreasing, with a doublet index. This suggest that these are sets of bound states separated by a finite size tunnelling barrier. We obtain sublattice polarization of the lowest doublet from  $P_{sub}=(\frac{dE}{dm_0})/N_{el}$, for the opposite mass field $P_{sub}(N_{el}=8)\sim 0.077$ and $P_{sub}(N_{el}=10)\sim 0.08$, which are comparable, and around three times smaller than $P_{sub}(N_{el}=9)\sim 0.25$.

\section{Filling factors between $\nu=+3$ and $\nu=+4$}
\begin{figure}
\begin{center}
\includegraphics[width=0.99\columnwidth]{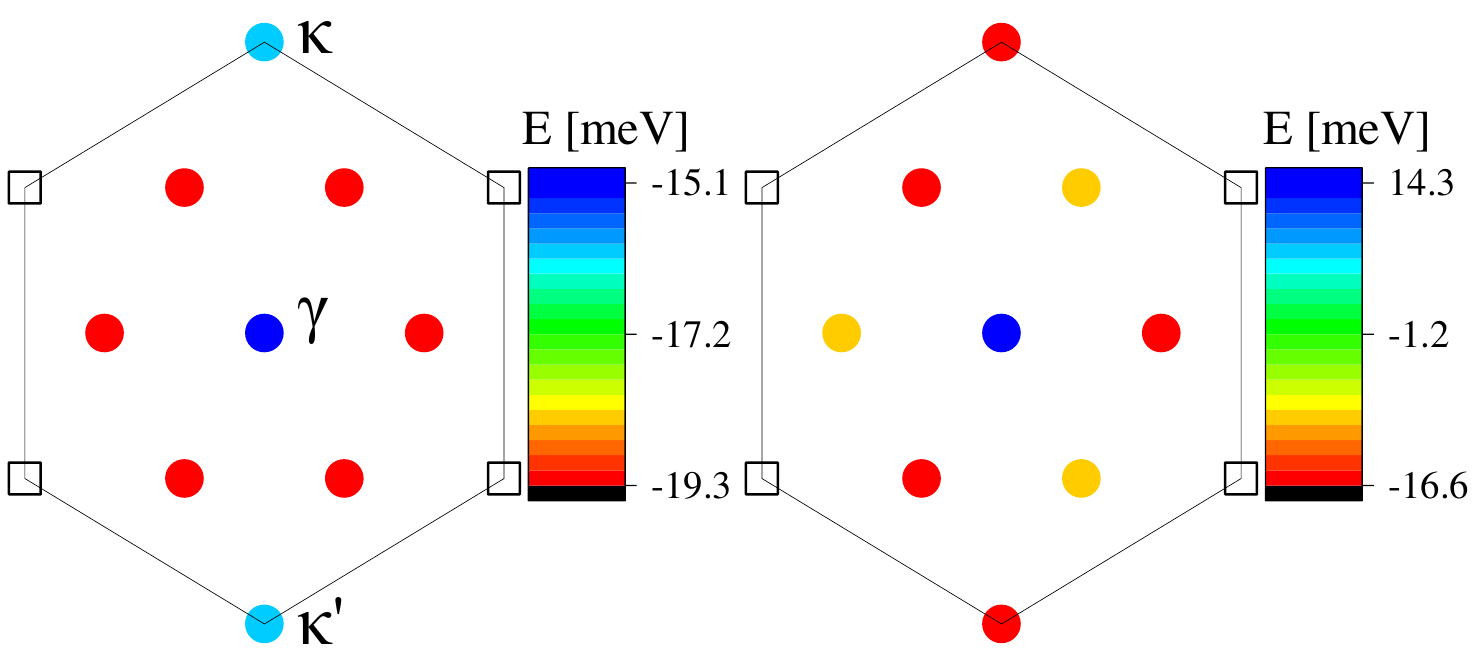}
\caption{Quasiparticle energies of full flat bands for 
$M=9$ ($N_x = 3$, $N_y = 3$) moir\'e unit cell systems at twist angle $\theta=1.1$,
interaction strength parameter $\epsilon^{-1}=0.05$, and inter-layer hopping parameters 
$\omega_1=110$ meV for inter sublattice tunneling and $\omega_0=0.8\omega_1$ for 
intrasublattice tunneling. The valence band is on the left and the conduction band on the right. 
Empty squares are used to mark momentum points that are equivalent to those included in the discrete mesh.}
\label{Fig:FiniteSizeBandsHoles}
\end{center}
\end{figure}

\begin{figure}
\begin{center}
\includegraphics[width=0.99\columnwidth]{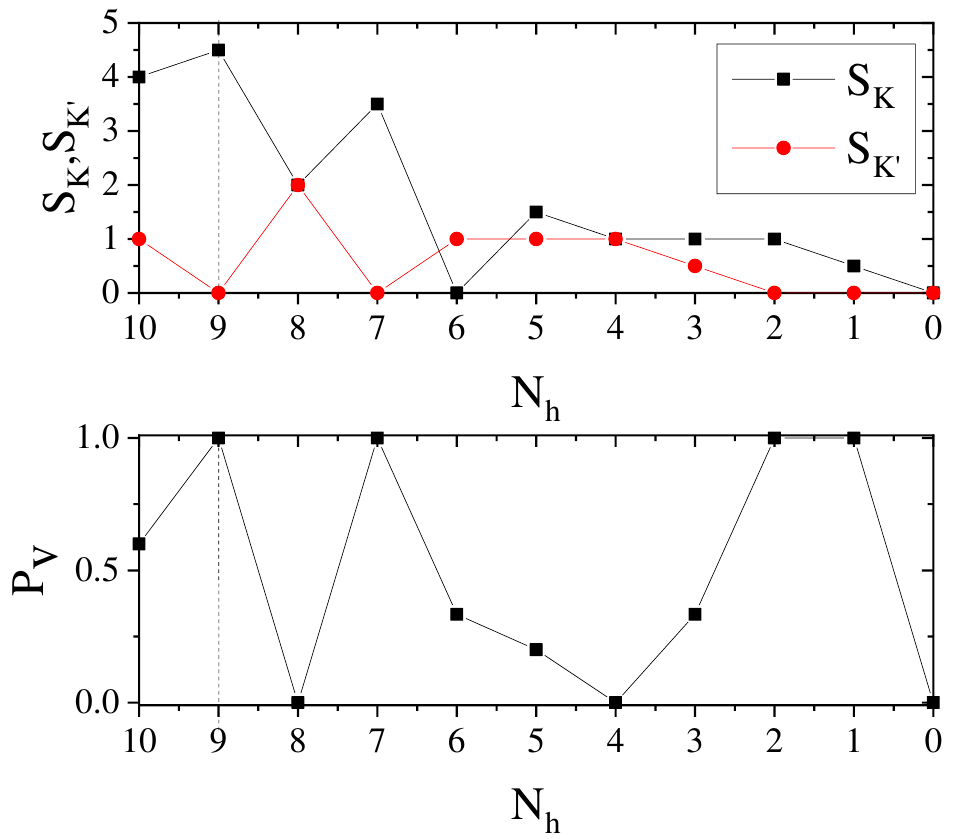}
\caption{The ground state properties as a function of a filling factor for holes. Top: Total spin $S_K$ and $S_{K'}$ in each valley. Bottom: Valley polarization $P_v$. The integer filling $\nu=+3$ is indicated by a dash line and corresponds to $N_{h}=9$.}
\label{Fig:PhaseHoles}
\end{center}
\end{figure}

We perform particle-hole transformation and investigate electronic properties in a vicinity of the full flat bands. The quasiparticle band dispersions for a single hole are shown in Fig. \ref{Fig:FiniteSizeBandsHoles} for the same set of parameters as in the case of empty bands considered in this work. The band dispersion is very similar to the one for electrons confirming particle-hole symmetry in the system. We depopulate these bands, or equivalently create holes, and analyze properties of the ground state between filling factors $\nu=+3$ and $\nu=+4$. The integer filling factor $\nu=+3$ is for a system with $N_h=9$ holes. The holes phase diagram is shown in Fig. \ref{Fig:PhaseHoles}. Similarities with the electron phase diagram are clearly visible. For low densities, no clear flavor polarization is observed with a transition to full spin polarization for $N_h > 6$. One can notice no full valley depolarization for $N_h = 10$, in contrast to the electron case. Summarizing, for a model considered in this work, the system has the electron-hole symmetry up to a very good approximation.

\section{Flat bands at Charge Neutrality}
\begin{figure}
\begin{center}
\includegraphics[width=0.99\columnwidth]{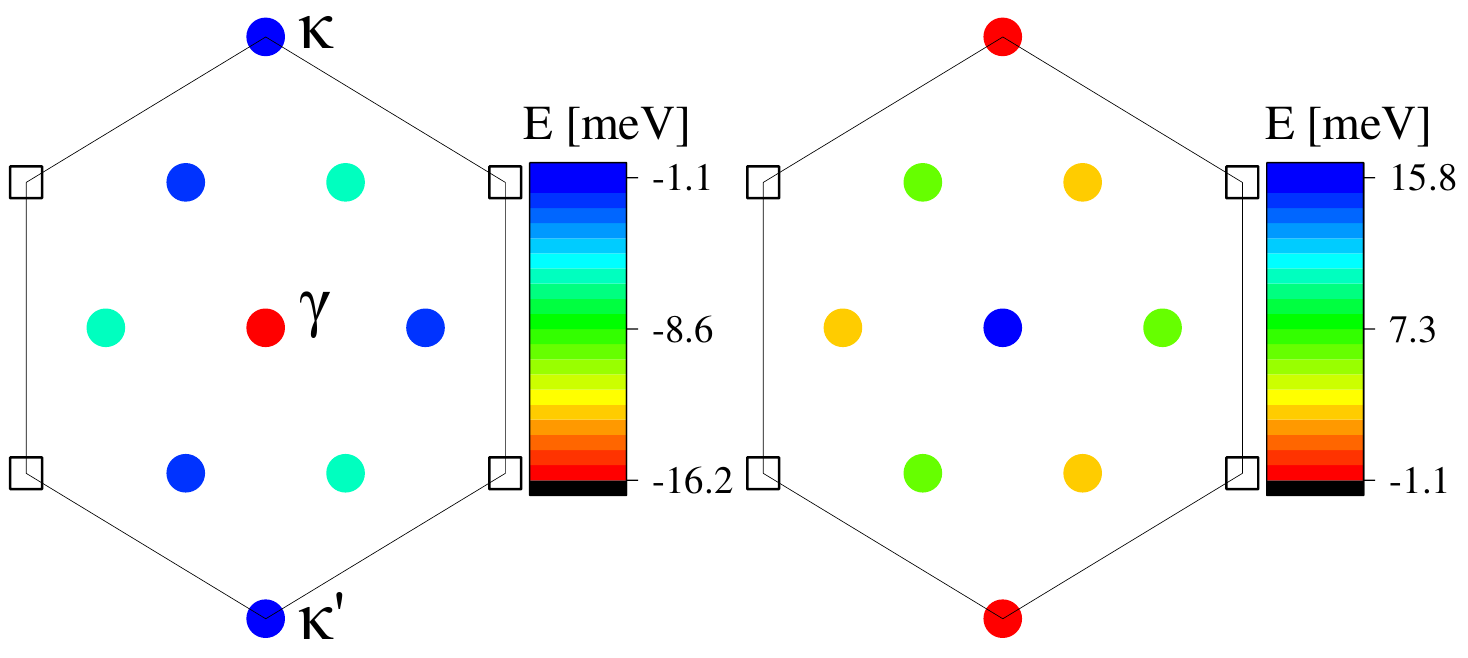}
\caption{Quasiparticle energies at charge neutrality for 
$M=9$ ($N_x = 3$, $N_y = 3$) moir\'e unit cell systems at twist angle $\theta=1.1$,
interaction strength parameter $eps^{-1}=0.05$, and inter-layer hopping parameters 
$\omega_1=110$ meV for inter sublattice tunneling and $\omega_0=0.8\omega_1$ for 
intrasublattice tunneling. The valence band is on the left and the conduction band on the right. 
Empty squares are used to mark momentum points that are equivalent to those included in the discrete mesh.}
\label{Fig:Fig_CN_bands}
\end{center}
\end{figure}

Fig. \ref{Fig:Fig_CN_bands} shows the quasi-particle bands at charge neutrality. 
In contrast to the empty flat band quasiparticles, 
the bottom of the valence band is at $\gamma$, as in the non-interacting model, the bands have approximate particle-hole symmetry, however the energy difference of two bands at $\gamma$ is significantly larger, around 30 meV, in contrast to few meV without the self-energies effect of remote bands.

\section{Many-body exact diagonalization calculations}
\subsection{Ground state quantum numbers estimation}
The total Hilbert space of four possible flavors and two, valence and conduction bands, in a translationally invariant system is divided into smaller subspaces, with a given number of electrons in each valley $N_{K}$ and $N_{K'}$, and labelled by quantum numbers: total spin $S_K$ and $S_{K'}$, azimuthal spin $S^z_{K}$ and $S^z_{K'}$, and total momentum $(K_x,K_y)$ in two directions determined by modulo of a number of unit cells $N_x$ and $N_y$ along two Moire lattice vectors. The basis is constructed in an occupation number configurations, distributing particles with a given spin $z$ azimuthal quantum numbers on Hartree-Fock quasi-particle $(k_x,k_y)$ momentum states. The many-body Hamiltonian is diagonalized in $S^z_{K}$ and $S^z_{K'}$ subspaces, and we do not rotate a Hamiltonian matrix to a $S_K$ and $S_{K'}$ basis as this is an additional computational cost, but rather determine the ground state total spins based on $S^z_{K}$ and $S^z_{K'}$ energy eigenvalues degeneracy. All of our exact diagonalization (ED) calculations are limited only by a maximal matrix size of a given subspace. The largest subspace corresponds to the lowest possible pair of $S^z_{K}$, $S^z_{K'}$, because they contain states with quantum numbers for all total spins $S_K$, $S_{K'}$. Using a notation $(N_{K\downarrow},N_{K\uparrow},N_{K'\downarrow},N_{K'\uparrow})$ we can distinguish possible particle distributions over four flavors: the first two particle numbers correspond to two spins in valley $K$ and the last two particle numbers to two possible spins in valley $K'$. 

We present a method of determining the ground state quantum numbers based on a system with $N_{el}=9$ particles, $\nu=-3$. We have five possible $N_{K}$ and $N_{K'}$ particle distributions with the lowest spin $z$ quantum numbers: $(4,5,0,0)$ with $N_{K}=9$ and $N_{K'}=0$, $(4,4,1,0)$ with $N_{K}=8$ and $N_{K'}=1$, $(4,3,1,1)$ with $N_{K}=7$ and $N_{K'}=2$, $(3,3,2,1)$ with $N_{K}=6$ and $N_{K'}=3$, $(3,2,2,2)$ with $N_{K}=5$ and $N_{K'}=4$, taking into account the fact that due to time-reversal and SU(2)$\times$SU(2) symmetries in the system, exchanging two valleys or two spins gives the subspace with identical energy spectra, e.g. $(4,5,0,0)$, $(5,4,0,0)$, $(0,0,4,5)$, $(0,0,5,4)$ have the same spectra, but energy spectra for $(4,5,0,0)$ and $(4,0,5,0)$ are in general different due to valley anisotropy energy. We can perform full ED calculations for $(4,5,0,0)$ and $(4,4,1,0)$ subspaces, but not for $(4,3,1,1)$, $(3,3,2,1)$, $(3,2,2,2)$, due to limited memory access and time of calculations. However, for higher spin $z$ subspaces full ED calculations are accessible, e.g. for a particle distribution with $N_{K}=7$ and $N_{K'}=2$ we can do full ED calculations for $(7,0,2,0)$, $(7,0,1,1)$, $(6,1,1,1)$, $(5,2,1,1)$, and determine an upper bound of total spins $S_K$ and $S_{K'}$ for the lowest eigenstate. This is estimated based on $S^z_{K}$ and $S^z_{K'}$ degeneracy forming a given $S_K$ and $S_{K'}$ multiplet. In this case, one can not be completely sure whether the total spins of the ground state are not contained in the subspace $N_{K}=7$ and $N_{K'}=2$ and correspond to $S_K = 0.5$ and $S_{K'} = 0$, due to lack of possibility of calculations for $(4,3,1,1)$, but typically energies of $S_K$ and $S_{K'}$ gradually changes with their magnitude. Indeed, we observe such a situation in most of the cases, the energies of the largest possible $S_K$ and $S_{K'}$ are usually the lowest, and when $S_K$ and $S_{K'}$ decrease, their energy increases up to the highest energy with the lowest $S_K$ and $S_{K'}$, suggesting that full spin polarization is quite robust for different fillings. 

\subsection{Truncated Hilbert subspace calculations}
\begin{figure}
\begin{center}
\includegraphics[width=0.99\columnwidth]{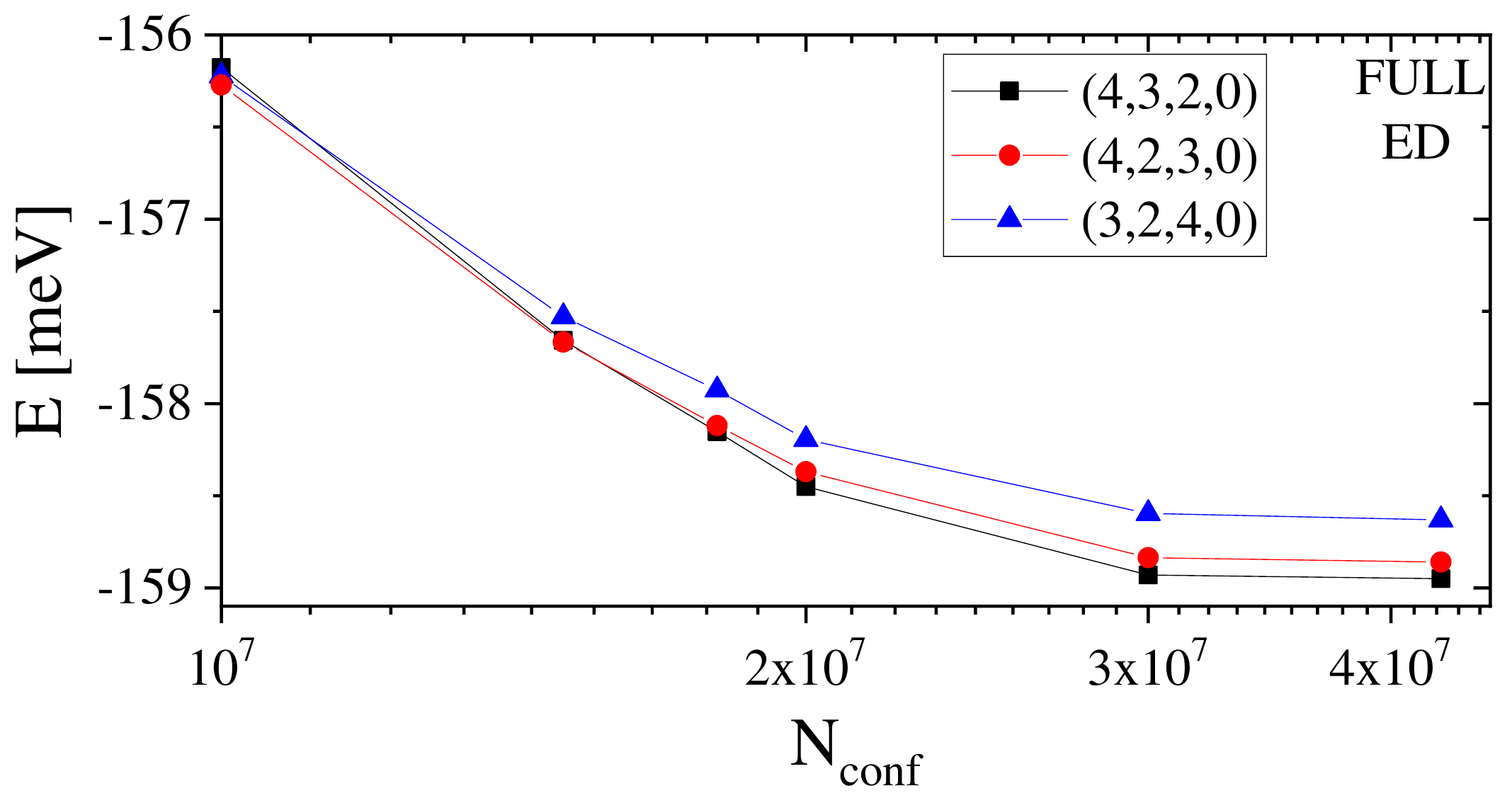}
\caption{The lowest energy eigenvalues as a function of a number of configurations $N_{conf}$ taken into truncated Hilbert space calculations for distributions of $N_{el}=9$. Top: $(4,3,2,0)$ with $P_v = 5/9$ (black squares), and $(4,2,3,0)$ with $P_v = 3/9$ (red circles), and $(3,2,4,0)$ with $P_v = 1/9$ (blue triangles).}
\label{Fig:SM_Nconf}
\end{center}
\end{figure}
The occupation number configurations can be sorted by their Hamiltonian expectation values and a truncated Hilbert space can be obtained by neglecting higher energetic ones. One can determine approximately valley polarization $P_v$ of the ground state by comparing the lowest energies of two or three different particle distributions for calculations in a truncated Hilbert space, analyzing results as a function of a truncated Hilbert space size. The eigenvalues within a given subspace converge monotonically with a number of configurations taken, giving always an upper bound of the energy of a given state. Tempo of convergence usually depends on a matrix size, but considering different particle distributions among valleys but with the same or similar number of particles within a given flavor, gives matrices of the same or similar size and a reliable comparison. While it may be difficult to reach a satisfactory approximation to the energy of the ground state by taking only a few percentage from a  total number of configurations, but based on the energy difference between the lowest states as a function of a truncated Hilbert space size, we can estimate which of particle distributions is favorable, and has a tendency to have the lower energy, as this difference usually remains constant or even increases. We can perform similar comparisons for pairs or triads of particle distributions getting approximate evaluation of valley polarization $P_v$. In Fig. \ref{Fig:SM_Nconf} we show energies as a function of truncated Hilbert space size for particle distributions $(4,3,2,0)$ ($P_v = 5/9$), $(4,2,3,0)$ ($P_v = 3/9$), $(3,2,4,0)$ ($P_v = 1/9$). In this case, we can reach the full Hilbert space size and one can clearly eliminate higher energetic particle distributions. This confirms that truncated Hilbert space calculations can give reliable information about the ground state properties. Here, the highest possible valley polarization is energetically favorable, $P_v = 5/9$ has the lowest energy. In general for all comparison that we performed for filling $\nu=-3$, the lowest energy in a given sector of valley polarization is always lower when compared with a smaller valley polarization sector. After determining valley polarization $P_v$, one can compare the lowest eigenvalues from different large-spin $z$ subspaces obtained from full ED and follow an estimation procedure described in a previous subsection in order to approximately determine total spins $S_K$ and $S_{K'}$. 

\subsection{Filling $\nu=-2$}
\begin{figure}[h]
\begin{center}
\includegraphics[width=0.99\columnwidth]{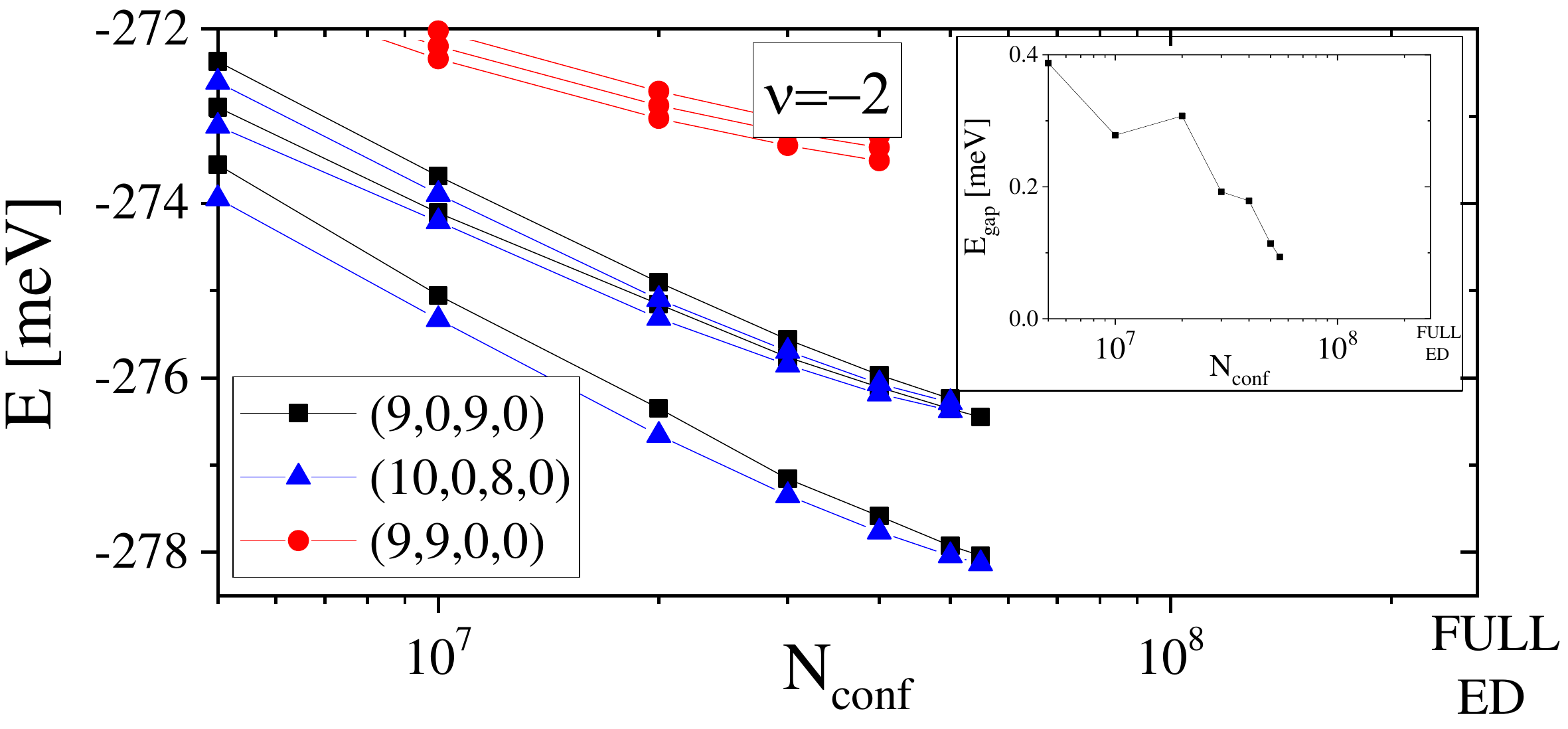}
\caption{Low energy spectrum as a function of a number of configurations $N_{conf}$ taken into truncated Hilbert space calculations for three different particle distributions of $N_{el}=18$, $(9,0,9,0)$ with $P_v = 0$ (black squares), and $(9,9,0,0)$ with $P_v = 1$ (red circles), $(10,0,8,0)$ with $P_v = 1/9$ (blue triangles). The inset shows the energy gap between the two lowest states.}
\label{Fig:SM_nu_min2}
\end{center}
\end{figure}

Using truncated Hilbert space calculations method described in the previous section, we analyze the nature of $\nu=-2$. Fig. \ref{Fig:SM_nu_min2} shows the low energy spectrum as a function of Hilbert space size given by a number of configurations taken $N_{conf}$ comparing three different particle distributions with the highest azimuthal quantum numbers $S^z_{K}$ and $S^z_{K'}$. Two lowest states correspond to fully valley unpolarized state $(9,0,9,0)$ and valley unpolarized state $(10,0,8,0)$ with intervalley excitation from the state $(9,0,9,0)$, when one of particles from equally distributed particles among two valley is moved to the other one creating slightly valley polarized state. The energies of the two lowest states are comparable with a decreasing energy between them with a truncated Hilbert space size (the inset), suggesting a crossing such that $(9,0,9,0)$ state is the ground state. 